\documentclass[Crown,sageh,times]{sagej}

\usepackage{moreverb,url}
\usepackage[colorlinks,bookmarksopen,bookmarksnumbered,citecolor=blue,urlcolor=blue]{hyperref}

\usepackage{graphicx}%
\usepackage{dcolumn}%
\usepackage{bm}%

\begin{document}

\title{Identifying sinks and sources of human flows: A new approach to characterizing urban structures}

\author{Takaaki Aoki\affilnum{1}, Shota Fujishima\affilnum{2}, and Naoya Fujiwara\affilnum{3-7}}
\affiliation{\affilnum{1}Graduate school of Data Science, Shiga University, Hikone, Japan\\
  \affilnum{2}Graduate School of Economics, Hitotsubashi University, Tokyo, Japan\\
  \affilnum{3}Graduate School of Information Sciences, Tohoku University, Sendai, Japan\\
  \affilnum{4}PRESTO, Japan Science and Technology Agency, Kawaguchi, Japan\\
  \affilnum{5}Center for Spatial Information Science, the University of Tokyo, Kashiwa, Japan\\
  \affilnum{6}Institute of Industrial Science, the University of Tokyo, Tokyo, Japan\\
  \affilnum{7}Tough Cyberphysical AI Research Center, Tohoku University, Sendai, Japan
}

\corrauth{Takaaki Aoki, Graduate school of Data Science, Shiga University, Hikone 522-8522, Japan.}
\email{takaaki.aoki.work@gmail.com}

\begin{abstract}
  Human flow data are rich behavioral data relevant to people's decision-making regarding where to live, work, go shopping, etc., and provide vital information for identifying city centers.
However, it is not as easy to understand massive relational data, and datasets have often been reduced merely to the statistics of trip counts at destinations,
discarding relational information from origin to destination.
In this study, we propose an alternative center identification method based on human mobility data.
This method extracts the scalar potential field of human trips based on combinatorial Hodge theory. It detects not only statistically significant attractive locations as the sinks of human trips but also significant origins as the sources of trips.
As a case study, we identify the sinks and sources of commuting and shopping trips in the Tokyo metropolitan area.
This aim-specific analysis leads to a combinatorial classification of city centers based on the distinct aspects of human mobility.
The proposed method can be applied to other mobility datasets with relevant properties and helps us examine the complex spatial structures in contemporary metropolitan areas from the multiple perspectives of human mobility.
\end{abstract}


\maketitle

\section{Introduction}
A city center indicates an area where commerce, entertainment, shopping, and political power are concentrated in high-rise buildings.
In formal terms, it often indicates a central business district (CBD) focusing on the commercial, business, and financial aspects, but this term could exclude other cultural, historical, and political factors.
From a classical viewpoint such as the von Th\"{u}nen model \citep{Thunen1966}, a city has a single center.
However, cities have become more complex and most metropolitan areas are polycentric, along with the development of transportation systems, residential architecture, and lifestyles in society.
Moreover, changes increased during the COVID-19 pandemic due to the drastic increase in the number of remote workers.

To reveal the urban spatial structure in a data-driven manner,
city centers have often been identified as having a higher concentration of employment than the surrounding suburbs.
However, this identification is not easy due to the arbitrary choice of threshold parameters and controversy about the quantitative definition of centers and subcenters, and several identification methods have been proposed to overcome the weaknesses of existing methods \citep{GIULIANO1991,McMillenSmith2003,Redfearn2007,Arribas-Bel2014}.
Moreover, highly granular and multiplex datasets, such as three-dimensional building datasets covering entire countries in Spain\footnote{\url{http://www.sedecatastro.gob.es/}} and Japan \footnote{\url{https://www.mlit.go.jp/plateau/}}, human mobility datasets based on mobile phone tracking \citep{Barbosa2018}, and community-based point-of-interest (POI) datasets \footnote{\url{https://foursquare.com/}}, have recently become available and have driven the development of relevant methods \citep{Duranton2021}.

In particular, human flow data are vital for urban spatial structures, being collected through person-trip surveys, and more recently, by mobile phone tracking.
These mobility data are aggregated as an \textit{origin-destination (OD) matrix} and describe the relationships between locations by showing how many people move from one location (origin) to another (destination) based on the type of trip (e.g., commuting trips from home to the workplace and other trips for shopping, entertainment, and schoolings).
These multiplex behavioral data of many people are the result of decision making on where people live, work, go shopping, and other miscellaneous activities, also involving the choices of transportation method and other factors.
These data have utilized to assess the connectivity among cities for functional policentricity \citep{Burger2012,Burger2014,Green2007}, thus
revisiting the center definition \citep{Carlos2013},
incorporating the diversity of human activities into a centrality index \citep{Zhong2017},
and detecting  clusters of venues \citep{Sun2016}.

\begin{figure*}
  \centering
  \includegraphics[width=\linewidth]{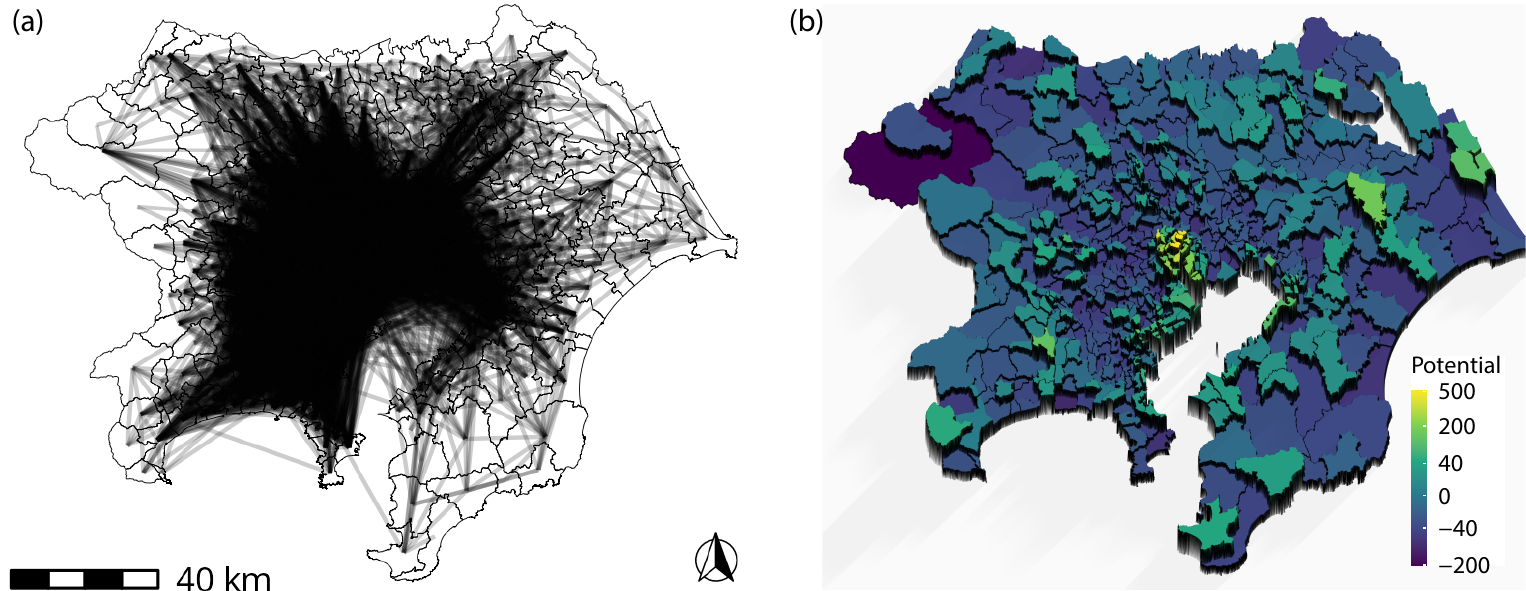}
  \caption{
    Visualizing the potential of human flows.
    (\textbf{a}) The links between the origins and destinations for the net movement of people by daily commuting trips in the Tokyo metropolitan area in 2018. 
    (\textbf{b}) 
    The potential field of the commuting flow shown in (a) by adopting the unique decomposition method by the combinatorial Hodge theory.
    The locations with higher potential attract commuters from those with lower potential.
  }
  \label{fig:potential_intro}
\end{figure*}

However, it is difficult to visualize the massive relational data on human flows.
Figure \ref{fig:potential_intro}(a) shows the daily commuting trip data for the Tokyo metropolitan area.
There are many links from origins to destinations, which makes it unclear to visualize urban structures.
To avoid such information overload, flow datasets have been reduced to statistics of trip counts at each destination in most previous works on functional relation analysis \citep{Burger2012,Burger2014,Green2007,Carlos2013,Zhong2017},
thereby discarding essential information on the relationship between origin and destination.

In this study, we propose an alternative center identification method based on human mobility data.
To this end, we first define the scalar potential field of human flow by adopting the unique decomposition of a given OD matrix according to combinatorial Hodge theory \citep{Jiang2011,Aoki2022} (see the Methods section for details).
Using a metaphor for water flowing from a higher to a lower place, the potential landscape provides an intuitive perspective of human flows and reveals the sinks and sources of the flow on a map, as shown in Figure \ref{fig:potential_intro}(b).
We then statistically identify the significant sinks and sources in comparison with a counterfactual null model.
The detected sinks indicate attractive locations for human flows, and we define them as flow-based city centers.

This new approach has several noteworthy features.
First, it allows us to classify city centers based on the aims of trips, such as commuting or shopping.
Trips of people from one location to another have specific purposes or aims.
Therefore, aim-specific centers can be identified by selecting mobility data based on aims.
Some centers attract workers, while others attract shoppers.
As such, aim-specific centers will reveal the different aspects of the centers in a city.
Moreover, the mobility dataset often includes other valuable properties, such as age, gender, ethnicity, and transportation methods.
These properties can provide distinct perspectives on the urban structure.
Second, the method identifies the sources of human flows, leading to significant locations in a city as the origins of human flows.
Combined with the aim-specific analysis, a place could be the source of a commuting flow and simultaneously be attractive to shoppers, which might be seen in some commuter towns.
This method classifies locations based on multiple aspects of human mobility and helps us to examine the complex spatial structures in contemporary metropolitan areas.

As a case study, we demonstrate the proposed method for center identification in the Tokyo metropolitan area.
There are several reasons why we focused on this area.
First, Tokyo is one of the largest metropolitan areas in the world in terms of population, area, and economy.
As of 2015, the Tokyo metropolitan area had a population of 37 million, covering an area of 14 thousand square kilometers.
The GDP of Tokyo was 1.6 trillion dollars measured by purchasing power parity \citep{Trujillo_Parilla_2016}.
Second, historical areas have been developed as political centers over several hundreds of years and have undergone many urban developments based on their previous buildings, leading to complex urban structures.
Third, well-organized person-trip datasets are available over several decades,
including the trip aim and other valuable information about each trip, as described later.

This study is thus related to several research topics across various disciplines.
First, it is relevant to the identification methods of city centers reviewed in the next section.
Second, human mobility represents relational information between locations and is recognized as network data embedded in space.
This spatially embedded network has been intensively studied in network science \citep{Barthelemy2011} and has also been utilized to detect the limits of cities \citep{amini_impact_2014,zhong_detecting_2014,huang2018,FujishimaFujiwara2020}.
Third, from the viewpoint of network analysis, the potential of combinatorial Hodge theory has been used for the analysis of gene regulatory networks \citep{Maehara2019,Qie2022}, neural networks \citep{Miura2015,Haruna2016}, inter-firm transaction networks \citep{Kichikawa2019,Fujiwara2021}, and the correlation network of macroeconomic indicators \citep{Iyetomi2020}.

The remainder of this paper is organized as follows.
The literature review section reviews the literature on center identification methods.
The Methods section introduces the proposed identification method.
The Data and Preprocessing section describes our data sources and treatments.
The Results section demonstrates the method, showing the identified centers classified by multiple aspects of human mobility in Tokyo.
The Conclusions section summarizes the findings and concludes the paper.

\section{Literature review}
\label{sec:previous_identification_methods} 
In this section, we briefly review previous methods for center identification to clarify the position of the proposed method.
See extensive related reviews \citep{Duranton2021,Yu2021} for further details.

Center identification methods can be classified into two distinct groups:
(1) methods based on the employment density analysis
and
(2) methods based on the analysis of functional relations.
This classification is analogous to the concepts of \textit{morphological} and \textit{functional} polycentricity proposed by \citet{Burger2012}, although these concepts were initially introduced in the discussions on the balance between cities, not city identification.

\subsection{Employment density analysis}
The first group focuses on employment density.
In monocentric models, the employment density peaks at the CBD and sharply decreases toward the periphery.
Researchers have extended this gradient view to polycentric situations and empirically examined the spatial changes in  employment density to detect centers and subcenters. 

\cite{GIULIANO1991} defined a center as ``a contiguous set of zones, each with density above some cutoff $\bar{D}$, that together have at least $\bar{E}$ total employment and for which all the immediately adjacent zones outside the subcenter have density below $\bar{D}$.''
In their research on Los Angeles, the cutoffs were ``10-10'' or ``20-20,'' at least 10 (20) jobs per acre and 10,000 (20,000) total employment.
This clustering method is intuitively simple and is widely adopted by modifying of the cutoffs \citep{SmallSong1994,Song1994,CerveriWu1997,BogardFerry1999,AndersonBogard2001,CoffeyShearmur2001,ShearmerCoffey2002,GuilianoRedfrean2007}.
\cite{MunzGarcia2010} proposed alternative cutoffs by which subcenters are zones with a higher density than the metropolitan average and include at least 1\% of metropolitan employment.
A criticism of the clustering method is that researchers can arbitrarily choose the number of subcenters by turning the cutoff parameters \citep{McMillenSmith2003}.
Another drawback is the lack of statistical tests for detected subcenters \citep{Redfearn2007}.

Another approach is the method of regression-based models, in which centers are defined as areas with significantly higher employment densities than their surroundings. McMillen and his collaborators \citep{McMillenMcDonald1997,McMillenMcDonald1998,McMillen2001,McMillenSmith2003} proposed nonparametric methods, which detects potential subcenters as outliers by analyzing the significant residuals in locally weighted regression (LOESS) and then test whether the candidates have a statistically significant effect on employment density.
\cite{Leslie2010} adopted kernel density estimation and \cite{Craig2001} used quantile splines. 
\cite{Redfearn2007} proposed a different nonparametric method, which detects potential centers as local maxima in the smoothed surface by LOESS and uses a heuristic algorithm to optimize the flexible boundary of centers by testing their significant difference in employment density by a bootstrap method.
These methods adopt a standard tricube kernel function for LOESS, which has a tunable kernel window parameter. It should be noted that the parameter selection can affect the results of these methods \citep{Lv2017}.

Finally, local indicators of spatial association (LISA) is an identification method \citep{Arribas-Bel2015} based on local Moran's $I$, which examines the spatial autocorrelation of employment density.
Locations with high--high (high density with high-density neighbors) and those with high--low (high density with low-density neighbors) become parts of city centers.

\subsection{Functional relations analysis}
The second group focuses on spatial interactions or the connectivity between locations.
This approach is related to network science and graph theory, although some terminologies, such as ``degree'' or ``centrality,'' do not always have the same meanings.

In the seminal studies \citep{Green2007,Burger2012}, 
the authors discussed the balance between cities. %
The importance of a city is assessed by incoming trips from other cities using inter-city trip data, while functional polycentricity is discussed by assessing whether incoming commuters are evenly distributed among cities.
\citet{Burger2014} extended the analysis to the multiplex networks between cities,
such as journey-to-school, shopping, leisure, intra-firm, and inter-firm relationships.

As a center identification method, \citet{Zhong2017} examined intra-city trip data for Singapore and introduced a compound index as location centrality, which consists of the diversity of various trip aims and the density of incoming trips.
The centers were detected by thresholding the index over a given value.
\citet{Veneri2013} defined a flow centrality ratio between the in- and out-degree of trip network with the indices of productive variety and directional dominance, and identified central municipalities in Milan and Rome by thresholding the indices over 1.
Under another approach, \citet{Carlos2013} decomposed employment in a zone into resident workers who live in the same zone and commuters from other zones, and proposed a new definition of the city center as a significantly high-density location for both types of employment.
\citet{Sun2016} studied ``check-in'' data, which represent the trips among venues.
The authors detected several clusters of venues using either of three clustering algorithms and identified city centers (up to two) by matching a landmark close to the centroid of venue clusters.

Although having little interaction with the above literature, there are some other studies on center identification that use network data (e.g., \cite{Beckmann1985spatial,Mazzoli2019b}). 
They transform OD data into a two-dimensional vector field and derive a potential function by appealing to standard vector calculus.
The local maxima of the potential function are identified as city centers.
Although our method is technically similar to theirs,
we use an extension of vector calculus that is applicable to graphs,
such that the network structure of the OD flows is preserved.
See \cite{Aoki2022} for a detailed comparison between  our method and that of \cite{Mazzoli2019b}.

\subsection{Positioning of the proposed method in literature}
Our method belongs to the second group of functional relations analysis.
The proposed method focuses on the flow of people from the origins to destinations and identifies not only the significant sinks but also the significant sources without discarding the relational information between them.
This is contrary to most previous methods in the second group, in which the origin side of the trips was often neglected.

Moreover, the index is based on a rigorous mathematical definition, as described in the Methods section.
The proposed method statistically detects significant centers, which has rarely been done among the methods of the second group.

From the regional viewpoint of the Tokyo metropolitan area, \cite{Li2018} analyzed the employment density distribution using sufficient fine-scale and region-wide data.
They adopted the LISA method in the first group of methodologies.
By contrast, we discuss centers based on human flows and use commuting as well as other trips.

Compared to the previous study \citep{Aoki2022},
this study introduces a new statistical center identification method 
and classifies locations by combination of commuting and shopping flows in the Tokyo metropolitan area.

\section{Methods: Center identification method by human flows}
\label{sec:our_methods} 
Here, we introduce an identification method for city centers based on the human mobility data provided by an OD matrix.
First, we mathematically introduce the scalar potential of net movements using the unique decomposition of combinatorial Hodge theory \citep{Jiang2011,Aoki2022}.
Second, we statistically identify the locations that are significantly attractive compared with the counterfactual model.

\subsection{Potential landscape of human flows by combinatorial Hodge theory}
Scalar potential, or simply potential, is a popular mathematical concept adopted in various scientific fields that provides an intuitive representation of the hidden abilities from which flows are generated.
We introduce such a potential for human flows on a graph (network) between locations according to combinatorial Hodge theory, beyond the conventional vector calculus for a continuous vector field.

First, we show the final formula for the potential $s_i$ at location $i$ given an OD matrix $M$
(see the Supplementary Material for a detailed derivation):
\begin{align}
s  = - \Delta_0^{\dagger} \text{div} Y. \label{eq:potential}
\end{align}
On the right-hand side, matrix $Y$ represents the net movement of people:
\begin{align}
  Y =  M - M^{\intercal}, \label{eq:netflow}
\end{align}
where $M$ is the given OD matrix representing the number of trips from origin $i$ to destination $j$ using elements $M_{ij}$. $M^{\intercal}$ denotes the transpose of $M$ and
$\text{div} Y$ is the divergence of $Y$, which describes the amount of flow entering and leaving location $i$,
\begin{align}
  \text{(div $Y$)($i$)} 
= \sum_{j \text{ s.t. $\{i,j\} \in E$}} Y_{ij}
= \sum_{j \text{ s.t. $\{i,j\} \in E$}} \left[ M_{ij} - M_{ji} \right].  \label{eq:divergence}
\end{align}
$E$ denotes the set of edges in an undirected graph $G(V,E)$.
In the context of this study,
vertices $V$ correspond to the locations and edges $E$ are the connections between them.
$\Delta_0^{\dagger}$ denotes the Moore-Penrose inverse of the graph Laplacian $\Delta_0$
\begin{align}
  \left[ \Delta_0 \right]_{ij} =  \begin{cases}
    d(i) \quad &\text{if $i = j$}\\
    -1   \quad &\text{if $\{i,j\} \in E$}\\
    0    \quad &\text{otherwise}
  \end{cases},
\end{align}
where $d(i)$ is the degree (number of connections) of vertex $i$.
The graph Laplacian can be viewed as a matrix form of the discretized Laplace operator on the graph
and $\Delta_0 s$ represents diffusion as a divergence of the gradient of $s$.

Potential $s$ provides the flow according to its gradient
\begin{align}
  (\text{grad}\, s)(i, j) &=  s_j-s_i \quad \text{$\{i,j\} \in E$} \label{grad}.
\end{align}
$s$ is also called a negative potential.
The higher potential $s_i$, the more location $i$ collects trips from others, and the locations in which the flow converges are called ``sinks.''
In contrast to sinks, the locations with lower potentials from which flow originated are called ``sources.''

Distances $d_{ij}$ between locations $i$ and $j$ crucially affect the connections between them.
Trips will be rare for long distances, even if the potentials at the endpoints differ.
On the other hand, no trips are often observed for short distances.
This implies that zero movements have two distinct situations.
To distinguish the nonexistence of movements based  on distance and other factors,
we assumed that a pair of locations is connected, $\{i,j\} \in E $, if its road distance $d_{ij}$ is within a threshold $\theta$. 
Threshold $\theta$ is determined based on a given dataset and is chosen to include 99\% of all trips (see the Data and Preprocessing section for details).

In the special case of a complete graph, in which all pairs of locations are connected,
potential $s$ can be simplified as
\begin{align}
s_i = - \frac{1}{N} \text{div} Y =  \frac{\sum_{j \neq i} M_{ji} }{N}- \frac{\sum_{j \neq i} M_{ij}}{N}, \label{eq:potential_complete}
\end{align}
where $N$ denotes the number of locations.
In this case, the potential represents the balance between incoming and outgoing fluxes.

The potential is simple, particularly in the complete graph case 
and readers might wonder if the combinatorial Hodge theory is needed to create the metric,
as it could be also defined on an ad-hoc basis.
However, the rigorous mathematical definition leads to the following its beneficial properties.

First, the potential of equation \eqref{eq:potential} is the solution of the following optimization problem:
\begin{align}
  \min_s \sum_{\{i,j\} \in E} \left[ \text{(grad $s$)}(i,j) - Y_{ij} \right]^2
  = 
  \min_s \sum_{\{i,j\} \in E} \left[ (s_j - s_i) - Y_{ij} \right]^2. \label{eq:optimization}
\end{align}
This  means that potential $s$ minimizes the residual sum of squares between its gradient flow and the given net flow
and enables us to interpret the estimation of potential as a regression.

Furthermore, the residuals in \eqref{eq:optimization} are unrelated components to qualifying the sinks and sources of a flow.
The residuals, $\bar{Y}_{ij} (= Y_{ij} - (s_j - s_i))$, belongs to the kernel of divergence by orthogonal decomposition in the combinatorial Hodge theory.
This indicates that the incoming and outgoing fluxes at each location are balanced: $\sum_j \bar{Y}_{ij} = \sum_j \bar{Y}_{ji}$. There is no sink or source for this divergence-free flow.
By contrast, potential $s$ produces an acyclic flow based on its gradient.
This means that potential $s$ is not just a metric of attractiveness at each location,
but is descriptive of the relationships among locations.

Next, potential $s$ is a linear map of given net flow $Y$.
There are several advantages from the viewpoint of data analysis.
For example, trip datasets may contain noise, caused by inevitable errors or difficulties in data acquisition.
When a given net flow $Y$ is given by $Y^{\text{signal}} + \epsilon Y^{\text{noise}}$ with a small $\epsilon$, the potential of the summed flow is equal to the sum of the potentials of each flow:
$s = s^{\text{signal}} + \epsilon s^{\text{noise}}$.
This guarantees that the perturbed potential is shifted slightly by the noise from the true potential without unexpected magnification of the noise.
In the case when the flow consists of two types of flows, such as commuting and shopping,
linearity allows us to divide the entire potential into the parts of individual types.
These properties do not generally hold when we consider another intuitive metric, such as the ratio of incoming and outgoing fluxes.

\subsection{Significant sinks and sources on the potential landscape}
Next, we statistically identify significant sinks or sources compared with a counterfactual model.

The mean of potential $\frac{1}{N}\sum_i s_i$ is zero and zones with positive $s_i$ tend to attract trips from others on average.
However, threshold $s = 0$ is not effective in selecting significant locations,
because the potential can be positive even when a flow is randomly generated.
Here, we select a particular set of zones as significant sinks or sources that are very unlikely to be observed by chance.

As a statistical null model, we introduce  counterfactual net flow $\hat{Y}$, which is randomly determined.
The actual human flow is not random: when location $i$ collects many trips from location $j$, it will also collect many trips from another location $k$.
The counterfactual flow $\hat{Y}$ destroys the correlations among the locations.
Specifically, as a null model, we adopt a \textit{link shuffling} model among the \textit{randomized reference models} of networks \citep{gauvin2020randomized}.
In the counterfactual flow $\hat{Y}$ in graph $G(V,E)$,
association between edge $\{i,j\}$ and flow $Y_{ij}$ are randomly shuffled
and the sign of $Y_{ij}$ is flipped with a probability of $1/2$.
In other words, the null model destroys the spatial structures in flow $Y$ by reallocating the location indices while remaining the set of absolute net movements, $\{|Y_{ij}|\}$, of the original $Y$.

Using the null model, we perform a permutation test with Monte Carlo simulation, as performed in spatial analysis \citep{de2018geospatial}.
We generate many samples $\{\hat{Y}\}$ and calculate potential $\hat{s}_i$ using equation \eqref{eq:potential} for each piece of  surrogate data $\hat{Y}$.
We then obtain sample distribution $P(\hat{s}_i)$ of the potential at location $i$ in the null model.
Notably, we performed permutation among edges $E$ in graph $G(V,E)$, while
the locations or location-based variables are often permutated in spatial analysis.
We obtained the sample distribution by generating 100,000 samples for the results shown below
\footnote{
In the special case of a complete graph, the distribution at any location can be approximated by a normal distribution for a large $N$:
$\hat{s} \sim \mathcal{N}\left(0, \sigma_{\hat{s}}\right)$, where $\sigma^2_{\hat{s}} = \frac{1}{N^3} \sum_{i} \sum_{j \neq i} Y_{ij}^2$.
See the Supplementary Material for a detailed derivation.
}.

Using the obtained potential distribution of the null model
we determine the significance of actual potential $s_i$.
For zones with positive potential, a multiple-comparison test is performed using the $p$-value of $P(\hat{s}_i > s_i)$ under the control of the false discovery rate (FDR), $\alpha$, with a standard Benjamini-Hochberg procedure \citep{BHMethod}.
We refer to the zones in which the null hypothesis is rejected as \textit{significant sinks}.
Similarly, for zones with negative potential, 
a multiple-comparison test is performed using the $p$-value of $P(\hat{s}_i < s_i)$ under FDR control,
and refer to the discovered zones as \textit{significant sources}.

\section{Data and Preprocessing}
\label{sec:data}
\subsection{Data}
We used datasets from the person-trip surveys conducted in 2018 in the Tokyo metropolitan area,
which are publicly available from the Tokyo Metropolitan Region Transportation Planning Commission.

The OD matrix denotes the number of trips aggregated by \textit{basic} geographical zones.
The \textit{basic} zone is defined by the Tokyo Metropolitan Region Transportation Planning Commission, and has typically tens of thousands of residents.
The dataset contains 615 zones.

The trips are categorized according by their aims.
For the trips from home to a destination, the aim types are
``commute,'' ``go to school,'' ``shopping and other private aims,'' ``sales,'' and ``others.''
We selected two major types by the number of trips: ``commute'' and ``shopping and other private aims,''
and refer to them as \textit{commuting} and \textit{shopping}, respectively, for simplicity.
There are 13 million commuting and 11 million shopping trips. 

\subsection{Estimation of threshold distance}
Given the Tokyo person-trip data, we calculate threshold distance $\theta$, above which trips are rarely observed.
Figure S2 shows the cumulative distribution of trip distance $P(X<d)$ as a function of trip distance $d$ for both commuting and shopping.
The trip distance is evaluated by the road distance between the centroids of the origin and destination zones, calculated by the Open Source Routing Machine (OSRM) based on Open Street Map \citep{OSRM}.
Threshold $\theta$ was determined based on criterion $P(X<\theta) = 0.99$. Ninety nine percent of the trips were observed to be within the threshold distance.
We obtained that $\theta$ is 65.3 km for commuting trips and 72.8 km for shopping trips. 

\subsection{Significance test}
We adopted FDR $\alpha$ = 0.05 in the following results.
For reference, the $p$-value of each geographical zone is summarized in the Supplementary Material (Figure S3). 

\begin{figure*}
  \centering
  \includegraphics[width=0.7 \linewidth]{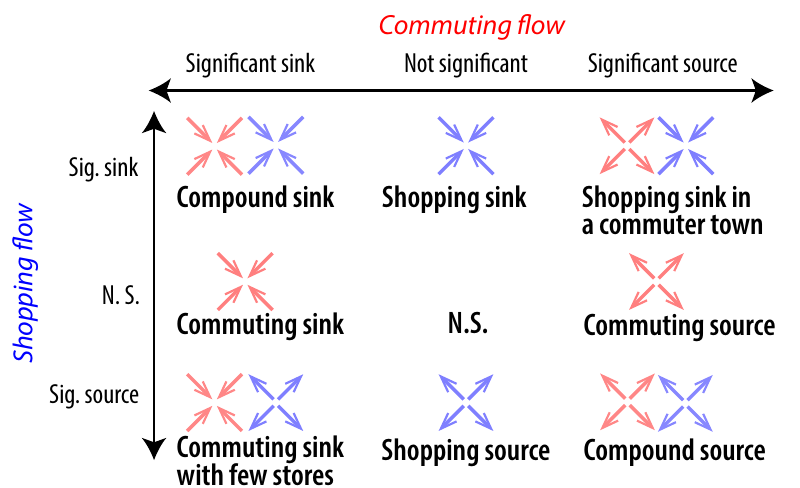}
  \caption{
    Combinatorial classification of significant locations by commuting and shopping flows.
  }
  \label{fig:method}
\end{figure*}
\section{Results}
We demonstrate the method for center identification in the Tokyo metropolitan area
as a case study using the person-trip survey in 2018.
As described in the previous section,
we select two major trips: \textit{commuting} and \textit{shopping}.
For each of these trips, we show the potential landscape (Figure \ref{fig:potential_intro}(b) for commuting trips and Figure S1 for shopping trips).
In both figures, there are several peaks in the potential landscapes,
which will be centers in the metropolitan area.
We statistically identify the significant centers as sinks of human flows
and the significant sources from which human flows originated.

\subsection{Classification of centers}

Figure \ref{fig:method} illustrates the classification of significant locations of the human flows.
For each trip's aim, 
this method classifies the locations into significant sinks, significant sources, and others (not significant locations).
Therefore, there are $3 \times 3$ classes of locations.

On the sink side, the \textit{compound sink} is a geographical zone that is a significant sink for both commuting and shopping trips.
It is an attractive area for diverse human activities.
Following the argument by \citet{Jacobs1969}, in urban studies, the center is regarded not only as a highly concentrated place but also as a place with diverse types of activities \citep{Zhong2017}.
Therefore, a compound sink is regarded as a center or subcenter in this classification framework.
\textit{Commuting sink} is the zone where that is a significant sink for commuting trips but not a significant one for shopping trips.
The location attracts many commuters from other zones, but not shoppers.
This implies that the zone would include offices or factories. %
\textit{Shopping sink} is the zone that is a significant sink only for shopping trips.
The implies that the zone would have shopping malls or stores. %

On the source side, the \textit{compound source} is the zone that is a significant source for both commuting and shopping trips.
Residents go outside for work and shopping, but do not stay in this zone.
Such a location would be a dense residential area or a commuter town with many houses and a few offices or shops. 
\textit{Commuting source} is the zone where residents go outside for work, but not shopping.
The location would includes several stores for residents.
\textit{Shopping source} is the opposite zone to the previous one.
Residents go outside for shopping, but not commuting.

A \textit{shopping sink in a commuter town} and a \textit{commuting sink with few stores} are complex zones with opposite features for commuting and shopping flows.
In a \textit{shopping sink in a commuter town} zone, residents go outside to work, whereas people come to the zone for shopping from other zones.
One possible situation is the case of a large suburban shopping mall within a commuter town.
\textit{Commuting sink with few stores} has the opposite condition. There would be a suburban industrial area.

The remaining type is the non-significant area for both flows.
It is noted that non-significance does not always indicate a few trips from and to the area.
The incoming and outgoing fluxes are balanced in such an area.

\subsection{Classified locations in Tokyo}
\begin{figure*}
  \centering
  \includegraphics[width=\linewidth]{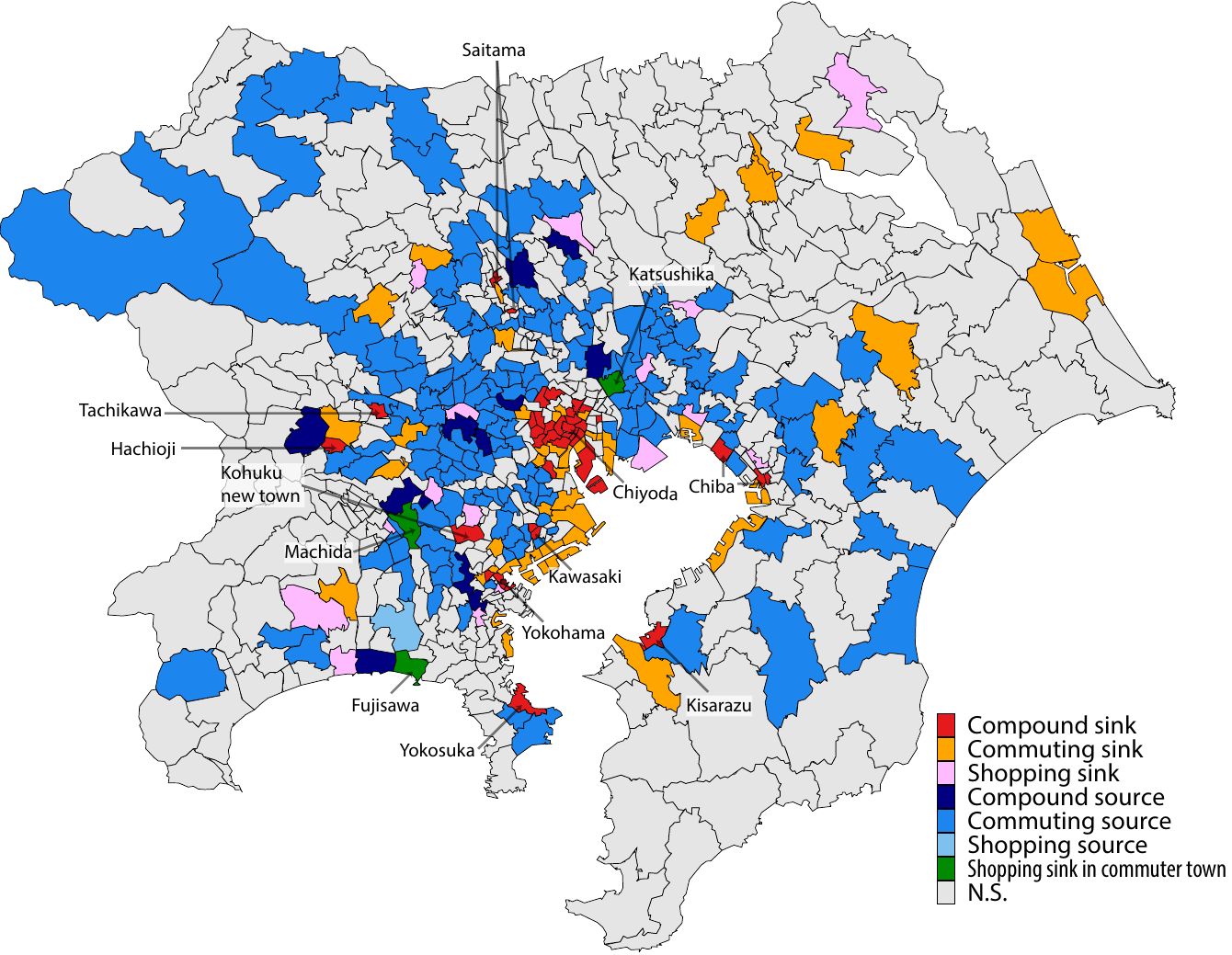}
  \caption{
    Classified zones in the Tokyo metropolitan area.
  }
  \label{fig:classification}
\end{figure*}
\begin{figure*}
  \centering
  \includegraphics[width=\linewidth]{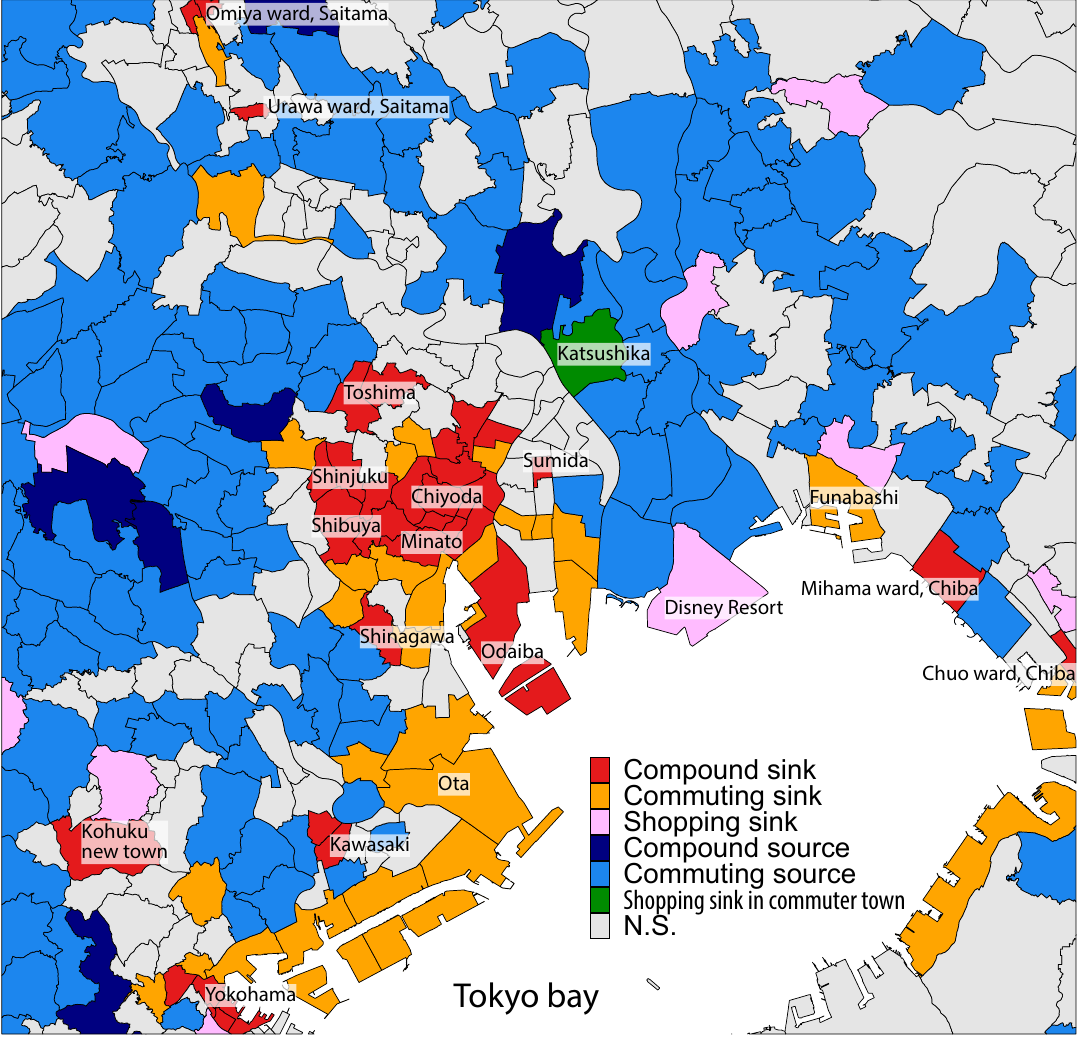}
  \caption{
    Classified zones in the central area in Tokyo.
  }
  \label{fig:classification_focus}
\end{figure*}

Figure \ref{fig:classification} shows the classified zones in the Tokyo metropolitan area
and reveals the spatial structure of the area.

Compound sinks are attractive places for commuting and shopping.
They were located in 
\textit{Chiyoda},
\textit{Kawasaki},
\textit{Yokohama},
\textit{Kohoku new town},
\textit{Tachikawa},
\textit{Hachioji},
\textit{Saitama},
\textit{Chiba},
\textit{Kisarazu},
and \textit{Yokosuka} areas.
These are well-known populated and attractive areas, listed in the business cores envisioned by the fourth and fifth National Capital Regional Development Plans of 1986 and 1999, as discussed later.

In particular, a cluster of contiguous compound sinks was seen around \textit{Chiyoda} city, which is known as the economic and political center of Japan.
It houses the headquarters of major enterprises, government institutions, and Tokyo Central Railway Station.
The cluster was extended to  neighboring cities \textit{Minato}, \textit{Chuo}, \textit{Shinjuku}, and \textit{Shibuya}, as shown in Figure \ref{fig:classification_focus}.
Another compound sink was seen in the \textit{Odaiba} area, a large artificial island on the Tokyo Bay side, which is known as an attractive location for shopping and tourism, not just business.
\textit{Toshima} was another compound sink, which has \textit{Ikebukuro} area known as one of the major attractive locations in Tokyo.

Many commuter sinks were observed in the Tokyo Bay area.
There are several major industrial areas, and these zones are regarded as major subcenters based on employment density analysis \citep{Li2018}.
Such industrial areas in the bay area are attractive to commuters but not to shoppers
and were identified as commuter sinks.
Moreover, the Tokyo Disney Resort at north of the Tokyo Bay area was identified as a shopping sink, as shown in Figure \ref{fig:classification_focus}.
Several commuter sinks were also identified, in addition to the bay area.
Most of them were located in business cores, as discussed later.
Shopping sinks were often seen in zones contiguous to sources.

Compound, commuting, and shopping sources were more broadly distributed than sinks.
Many sources surround compound sinks as the hinterlands of sources of human flows.

The \textit{shopping sink in a commuter town} type
was identified in
\textit{Machida},
\textit{Katsushika}, and 
\textit{Fujisawa} cities.
These zones attract shoppers from other zones and simultaneously serve as sources of commuters. 
These areas are commuter towns with shopping and tourist spots. 
By contrast, no zone was identified as \textit{commuting sink with few stores}.

\subsection{Validation}
\begin{figure*}
  \centering
  \includegraphics[width=\linewidth]{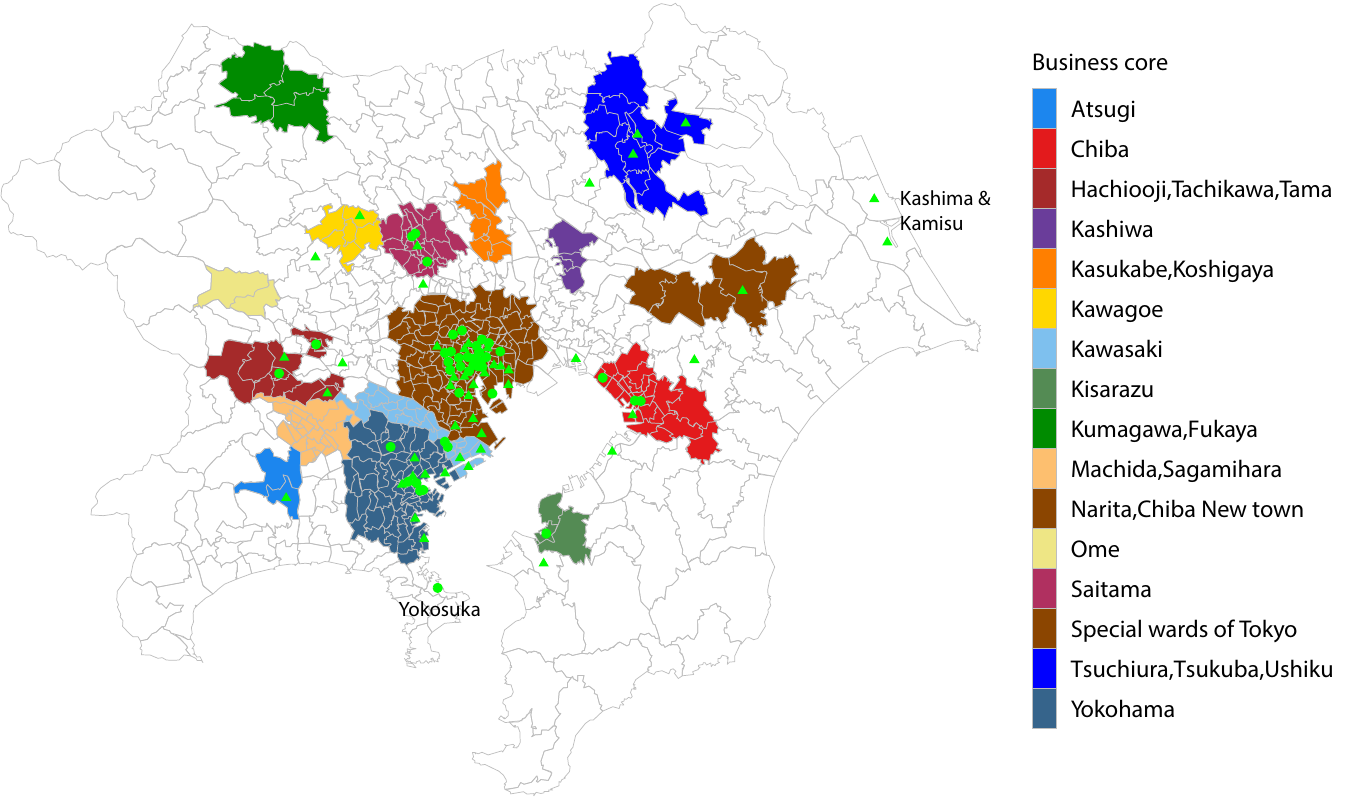}
  \caption{
    Comparing the detected \textit{compound sinks} by the proposed method with the business cores defined by the National Capital Regional Development Plan in 1999.
    The plan envisioned multi-nucleated urban structures to solve the over-concentration in special wards of Tokyo.
    The green points indicate the centroids of the detected compound sinks (circles) and commuting sinks (triangles).
  }
  \label{fig:bizcore}
\end{figure*}
It is inherently difficult to validate the identified centers because there is no ground truth for comparison.
However, we consider the business cores defined by the National Capital Regional Development Plans as a proxy for the ground truth, 
\cite{Li2018} also compared identified centers with business cores.
Furthermore, we introduce a score to evaluate matching with business cores and statistically test the comparison.

Business cores are the subcenters introduced in the Fourth National Capital Regional Development Plan of 1986,
which aimed to create multi-nucleated urban structures to solve several urban issues, such as over-concentration, traffic congestion, and high land prices in the special wards of Tokyo (\textit{tokubetsu-ku} in Japanese).
Eleven cores were selected in 1986 and extended to 15 cores in 1999 according to the Fifth National Capital Regional Development Plan.

There are several issues when considering business cores as centers.
First, the cores are political plans and are not guaranteed to be actual centers.
Second, some areas may have spontaneously developed into centers, unrelated to the plans.
Third, a business core corresponds to a city or a composite of cities. Their administrative boundaries are often too wide compared with the actual central geographical zones inside boundaries.
Fourth, the political plans also promoted facilities for recreation and shopping in addition to providing business support. This means that the envisioned center is not just for commuters but also for other human activities.
Concerning the first and second points, the following comparison between our results and the business cores could be regarded as an evaluation of the national plans by the person-trip data of 2018.

Figure \ref{fig:bizcore} shows the business cores and special wards of Tokyo, with the identified compound and commuting sinks.
All compound sinks were located in the special wards of Tokyo or either of the business cores, except for \textit{Yokosuka} city.
The commuting sinks have a similar tendency.
Most of them were located inside the business cores,
or in the zone adjacent zone to business cores,
except for the cities of \textit{Kashima} and \text{Kamitsu}.
These exceptional sinks might have developed unrelated to the National Development Plans of 1988 and 1999.
On the other hand, some business cores had no compound or commuting sinks, such as 
\textit{Ome}, \textit{Kasukabe \& Koshigaya}, \textit{Kashiwa}, \textit{Kumagaya \& Fukaya}, and \textit{Machida \&Sagamihara}.
This suggests that these areas have not developed as expected from the perspective of human flows.

Next, we define a matching score to evaluate how well the identified compound sinks match business cores.
We denote by $A$ the total area, except for the special wards of Tokyo and by $B$ the united area of all business cores.
The matching score is given by, 
\begin{align}
  \text{matching score} = \frac{\text{number of compound sinks in $B$}}{\text{total number of compound sinks in $A$}}.
\end{align}
The matching score was 0.94.
To test this value statistically, we consider a counterfactual model in which the sinks are randomly relocated within the area of $A$, while maintaining their total number. From the 100,000 observations in this null model, we found that the matching score is significantly high in one-sided tests, with a significance level of 0.01\%.
This result indicates that the detected compound sinks are located well within business cores at a rate that is much higher than chance. In other words, the identification method is validated regarding the relationships with business cores in the Tokyo metropolitan area.

\subsection{Comparison with previous methods}
\begin{figure*}
  \centering
  \includegraphics[width=12cm]{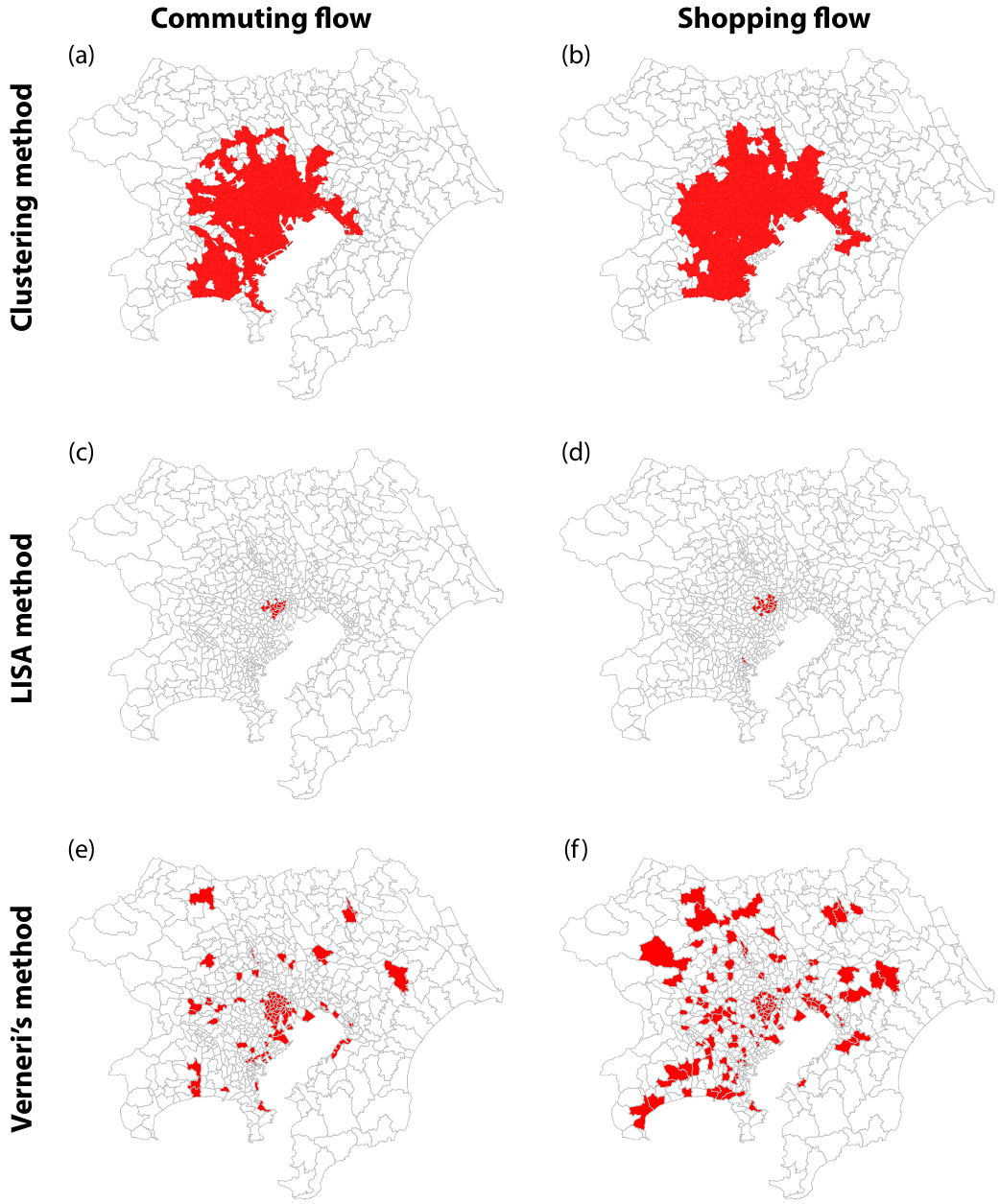}
  \caption{
Identified centers by other center identification methods
for commuting and shopping flow datasets in the Tokyo metropolitan area.
    (\textbf{a}) 
    The identified center by clustering method for the incoming trip density distribution of the commuting flows.
    (\textbf{b}) 
    The same as in (a), but for the shopping flow.
    (\textbf{c}) 
    The identified center by the LISA method for the incoming trip density of the commuting flow.
    (\textbf{d}) 
    The same as in (c), but for the shopping flow.
    (\textbf{e}) 
    The identified center by the method proposed by \cite{Veneri2013} for the commuting flow.
    (\textbf{f}) 
    The same as in (e), but for the shopping flow.
  }
  \label{fig:comparison}
\end{figure*}
Readers might wonder which locations are detected as centers if previous methods are applied to the same trip dataset and how the detected centers differ from those of the proposed method.

As discussed in the literature review section, 
incoming trip distribution at destinations has often been examined in the framework of functional relations \citep{Burger2012,Burger2014,Zhong2017,Carlos2013}.
Therefore, we mainly analyzed this metric for the commuting and shopping flow datasets in the Tokyo metropolitan area.
The details of the applied methods are summarized in the Supplementary Material.

First, we applied the clustering method as one of the standard identification methods.
In the literature, \citet{Zhong2017} used a similar cutoff method.
By using the automatic cutoff threshold proposed by \citet{MunzGarcia2010},
the clustering method detected a single giant cluster over several provinces as shown in Figure \ref{fig:comparison}(a,b).
Such a giant cluster using the clustering method was reported by \citet{Li2018} 
for employment density distribution in Tokyo and the authors adopted the LISA method instead.

Second, we applied the LISA method and found a single cluster in the central area of Tokyo without subcenters for commuting trips, with FDR $\alpha$ = 5\% (Figure \ref{fig:comparison}(c)). For shopping trips, a small area in the \textit{Yokohama} area was additionally detected (Figure \ref{fig:comparison}(d)).

Third, \citet{Veneri2013} proposed another definition of city centers and subcenters by commuting flow $M_{ij}$.
He argued that ``second-order subcenters'' are defined as locations that satisfy the following criteria:
(1) the flow centrality ratio ($FC \equiv \sum_j M_{ji} / \sum_{j} M_{ij}$) is greater than 1; and (2) the directional dominance index ($DII \equiv I_i / \bar{I}$) is greater than 1, where $I_i = \sum_j M_{ji}$ and $\bar{I}$ is the mean of $I_i$ over all zones.
Figure \ref{fig:comparison}(e) shows the second-order subcenters for commuting trips.
It should be noted that \citet{Veneri2013} imposed an additional condition on the diversity of activity to detect ``first-order subcenters,'' since the diversity is essential for centers \citep{Jacobs1969}.
We show the second-order subcenters for shopping trips in Figure \ref{fig:comparison}(f).

We found that the condition on flow centrality ($FC$) can be rewritten using the potential in the special case of a complete graph in Equation \eqref{eq:potential_complete}.
This allows us to impose a statistical test on the first condition by detecting significant sinks in the proposed framework.
The second condition on the directional dominance index is also rewritten by in-strength centrality in network analysis, and its outliers could be detected using the same framework of link shuffling that we adopted.
On the other hand, spatial interactions are often incomplete.
Thus, it is preferable to use the potential in a general graph given by Equation \eqref{eq:potential} than flow centrality $FC$.

In comparison to the proposed method by which several subcenters were detected as compound sinks,
the clustering method detected a single giant cluster spread across several provinces, and the LISA method detected a monocentric cluster in the central area of Tokyo for commuting trips.
The second-order subcenters defined by \citet{Veneri2013} were identified for similar zones as the significant sinks in the case of the commuting trips while they are not as similar in the case of shopping trips.

\section{Conclusions}
In this study, we proposed a center identification method based on human mobility data provided by an OD matrix.
The method extracts the scalar potential field of human trips based on combinatorial Hodge theory and detects not only statistically significant attractive locations as the sinks of human trips, but also significant origins as the sources of trips.

As a case study, using person-trip data from the Tokyo metropolitan area, we identified aim-specific sinks and sources by selecting trips by two major aims, commuting and shopping, and classified the locations into three $\times$ three types.
Compound sinks --- attractive locations for multiple human activities --- were detected in \textit{Chiyoda} and its surroundings and in other well-known attractive areas.
Almost all compound sinks are located in the business cores of the National Capital Regional Development Plans, whereas the detected sink in \textit{Yokosuka} city is located outside the business cores.
The aim-specific centers distinguish central areas.
For example, among the populated areas in the Tokyo Bay area, major industrial areas were detected as commuting sinks.
The geographical zone of the Tokyo Disney Resort was identified as a shopping sink.
Significant sources were broadly distributed around the sinks and formed the hinterlands in the metropolitan area.
Furthermore, we found a complex type of significant location --- a source of commuting flow and, simultaneously, a sink for shoppers.

This aim-specific classification reveals not only the locations of centers and subcenters in the Tokyo metropolitan area but also the diverse functional types of the detected zones based on people's movements.
For future work, the classification could be more specified using multi-faceted mobility datasets including factors such as age, gender, ethnicity, and transportation method.
For example, some zones attract young commuters, and other zones attract older adult shoppers.
Some shopping zones could attract one specific ethnicity or a diverse group of ethnicities.
Therefore, this method allows us to classify the significant zones based on multiple characteristics of human activities and helps us look into the complex spatial structures in contemporary metropolitan areas.

Finally, we discuss another approach to manage the massive relational information of a given OD matrix and the related research avenue for future work.
\citet{holmes1977graph} proposed a method for identifying the significant elements of the matrix.
This approach is relevant to the backbone methods in network science, which detects significant links of a given network dataset \citep{serrano2009extracting, tumminello2011statistically, li2014statistically,gemmetto2017irreducible, marcaccioli2019polya}.
The clustering of links is also helpful for finding informative patterns \citep{tao2016spatial, tao2017flowhdbscan}.
For future work, it will be essential to study pattern detection and the classification of significant connections between locations in a given OD matrix.

\begin{acks}
  We would like to thank T. Kobayashi, T. Ishihara, D. Zusai, and T. Akamatsu for the fruitful discussions.
\end{acks}
\begin{funding}
  This work was supported by the Research Institute for Mathematical Sciences, a joint research center of Kyoto University (NF)
  under JSPS KAKENHI Grant Number JP18K12776 and 22K18525 (SF) and JSPS KAKENHI Grant Number JP21H03507 (NF).
\end{funding}

%


\end{document}


\maketitle
\tableofcontents

\clearpage

\section{Potential landscape of human flow by combinatorial Hodge theory}
We begin with the potential of conventional vector calculus.
For a continuous vector field, scalar potential $\phi$ is given by Helmholtz's decomposition.
A smooth vector field $\vec{F}$ is uniquely decomposed into the gradient of scalar potential $\phi$ and curl of vector potential $\vec{A}$: $\vec{F} = -\nabla \phi + \nabla \times \vec{A}$.
Potential $\phi$ is obtained by solving the normal equation $ - \text{div}\, \vec{F} =  \nabla \cdot \left[ \nabla \phi - \nabla \cdot (\nabla \times \vec{A}) \right] = \nabla^2 \phi =  \Delta \phi$.

Next, this decomposition is extended to flow on a graph because a human flow is not given by a vector field but by a flow on a graph (network) between locations.
Let $G=(V,E)$ be an undirected, unweighted graph and $A_{ij}$ be a skew-symmetric matrix that represents the flow on the edge from $i$ to $j$.
The operators in vector calculus, grad, curl, and div are defined on a graph as follows:
\begin{align}
  (\text{grad}\, s)(i, j) &=  s_j-s_i \quad \text{for $\{i,j\} \in E$} \label{grad}, \\
(\text{curl}\, A)(i, j, k) &= A_{ij} + A_{jk}+A_{ki}\quad \text{for  $\{i,j,k\}$}: \{i, j\}, \{j, k\}, \{k, i\} \in E\  \label{curl},\\
  (\text{div}A)(i) &= \sum_{j \text{ s.t. } \{i,j\} \in E} A_{ij},
\end{align}
where $s$ denotes the potential function to be introduced.
The space of flow matrix $\mathcal{A}$ is orthogonally decomposed into the images and kernels of these operators as follows:
\begin{equation} 
  \mathcal{A}  = \text{im}(\text{grad})  \oplus \text{ker}(\Delta_1) \oplus  \text{im}(\text{curl}^*), \label{eq:decomposition}
\end{equation}
where ker($\Delta_1$) = ker(curl) $\cap$ ker(div) and $\text{curl}^*$ is the adjoint operator of the curl.
Potential $s$ can be obtained by projecting $A$ onto im(grad), owing to the orthogonal decomposition of \eqref{eq:decomposition},
With a Euclidean inner product in space $\mathcal{A}$, $\langle X,Y\rangle = \sum_{ \{i,j\} \in E} X_{ij}Y_{ij}$, the normal equation is given by
\begin{align}
  \Delta_0 s  = - \text{div} A,  \label{eq:normal_equation}
\end{align}
where $\Delta_0$ is the graph Laplacian
\begin{align}
  \left[ \Delta_0 \right]_{ij} =  \begin{cases}
    d(i) \quad &\text{if $i = j$}\\
    -1   \quad &\text{if $\{i,j\} \in E$}\\
    0    \quad &\text{otherwise}
  \end{cases},
\end{align}
where $d(i)$ is the degree of vertex $i$.
Finally, potential $s$ is expressed as the minimal-norm solution of \eqref{eq:normal_equation}:
\begin{align}
  s  = - \Delta_0^{\dagger} \text{div} A, \label{eq:potential}
\end{align}
where $\dagger$ denotes the Moore-Penrose inverse.

In this study,
vertices $V$ correspond to the locations given in the OD matrix and edges $E$ are the connections between them.
The skew-symmetric flow on the graph corresponds the net movement of people between locations, given by
\begin{align}
  Y =  M - M^{\intercal}, \label{eq:netflow}
\end{align}
where $M$ is the OD matrix in a given dataset representing the number of trips from origin $i$ to destination $j$ by its elements $M_{ij}$. $M^{\intercal}$ denotes the transpose of $M$.
$Y$ is skew-symmetric, $Y_{ij} = - Y_{ji}$, to which we can consider potential $s$.
According to Equation \eqref{eq:potential}, 
the potential for net human flow $Y$ is given by
\begin{align}
  s  = - \Delta_0^{\dagger} \text{div} Y. \label{eq:potential_netflow}
\end{align}

\clearpage

\section{Potential landscape in Tokyo metropolitan area}
\begin{figure}[h]
  \centering
  \includegraphics[]{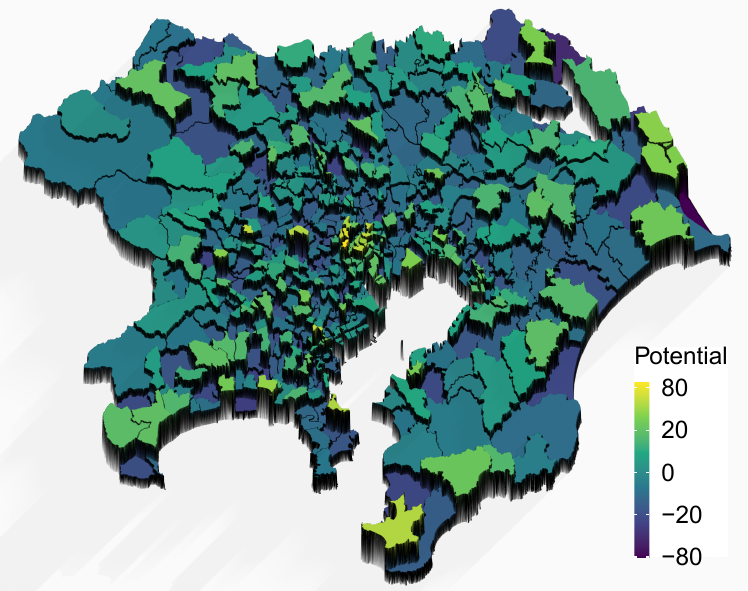}
  \caption{Potential landscapes for shopping trips.}
  \label{fig:potential_shopping}
\end{figure}

Figure \ref{fig:potential_shopping} shows the potential landscape for shopping trips in the 2018 person-trip survey data for the Tokyo metropolitan area (see Figure 1 for commuting trips).

\clearpage
\section{Threshold distance}
\begin{figure}[htb]
  \includegraphics[width=\linewidth]{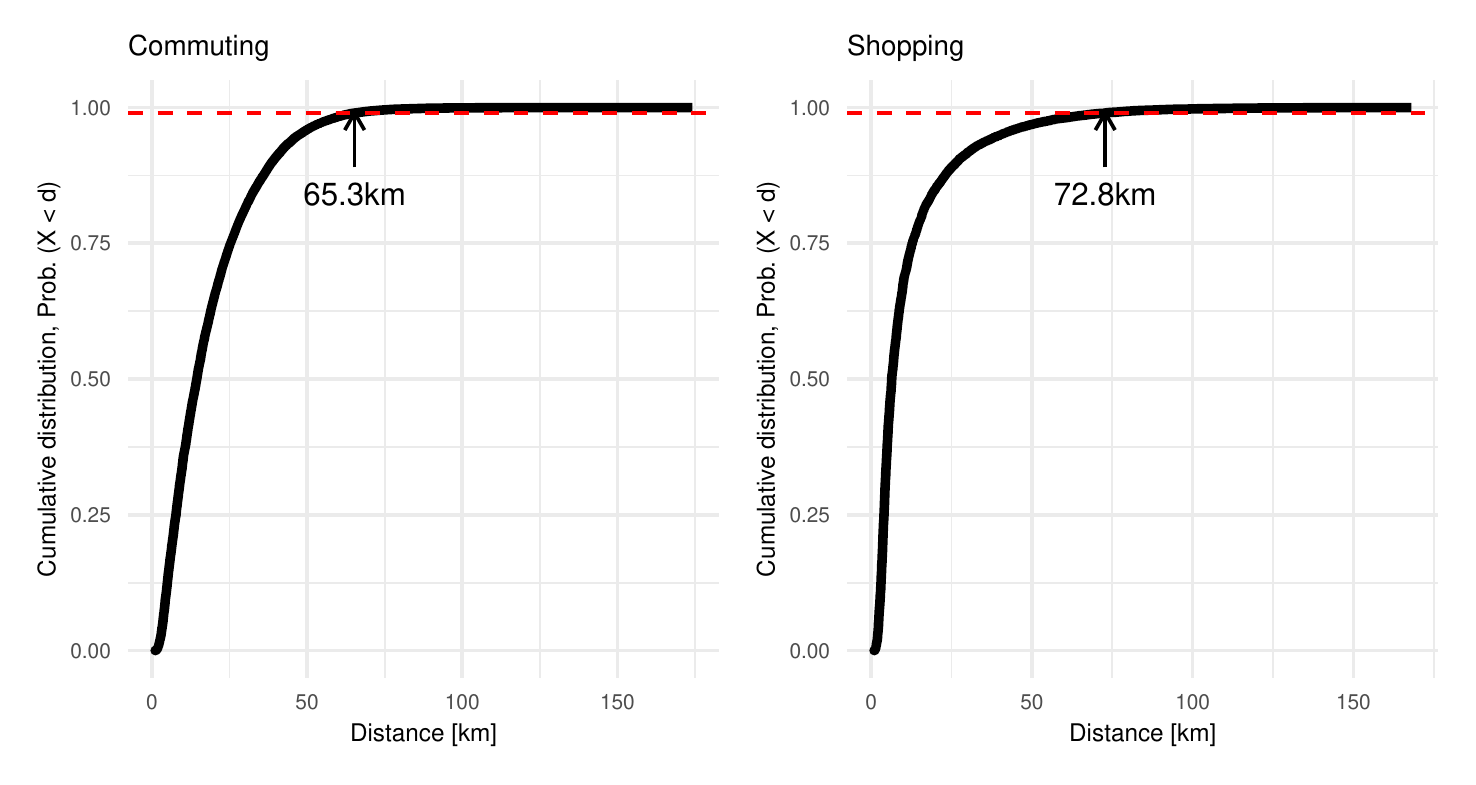}
  \caption{Cumulative distribution of trips, $P(X<d)$, as a function of trip distance $d$ in both cases of commuting and shopping. The horizontal dot line indicates $P(X<\theta)$ = 0.99.}
  \label{fig:distance}
\end{figure}

Given the Tokyo person-trip data, we calculate the threshold distance $\theta$, above which trips are rarely observed.

Figure \ref{fig:distance} shows the cumulative distribution of trips, $P(X<d)$, as a function of trip distance $d$ for both commuting and shopping.
The trip distance is evaluated by the road distance between the centroids of the origin and destination zones, calculated by the Open Source Routing Machine (OSRM) based on Open Street Map \citep{OSRM}.
Threshold $\theta$ is determined based on criterion $P(X<\theta) = 0.99$. Ninety-nine percent of the trips were within the threshold distance.
We obtained that $\theta$ = 65.3 km for commuting trips and 72.8 km for shopping trips.

\clearpage
\section{Potential distribution of the null model in the case of the complete graph}
In the special case that $G(V,E)$ is a complete graph,
potential $s$ is given by
\begin{align}
  s_i = - \frac{1}{N} \sum_{j \neq i} Y_{ij}, \label{eq:potential_without_distance}
 \end{align}
where $N$ denotes the number of locations.

In the counterfactual null model, elements of original matrix $Y$ are randomly relocated in their matrix indices, maintaining skew-symmetry.
Let us consider a random variable $X$ that is uniformly drawn from a set $\{Y_{ij} \mid i \neq j\}$
and define an approximated potential $\hat{s}$ of the null model:
\begin{align}
  \hat{s} =  - \frac{1}{N} \sum_{i = 1}^{N-1} X_{i},  \label{eq:potential_null}
\end{align}
where $\{ X_1, X_2, \cdots, X_{N-1}\}$ is a sequence of independent and identically distributed (i.i.d) random variables from $X$.
It should be noted that this sequence may contain duplicated elements of $Y$ because the elements are chosen with replacement, whereas the elements of $Y$ are chosen without replacement for a surrogate matrix $\hat{Y}$ in the original null model. Moreover, we do not maintain the skew-symmetry of $\hat{Y}$ here, contrary to the original null model.

This approximation of the null model allows us to explicitly describe the potential of the null model.
By using $\bar{X}_N \equiv \frac{1}{N} \sum_i^N X_i$, equation \eqref{eq:potential_null} can be rewritten as
\begin{align}
  \hat{s} = - \frac{N-1}{N} \bar{X}_{N-1}. \label{eq:potential_null_rewrite}
\end{align}
According to the central limit theorem, the distribution of $\bar{X}_{N-1}$ can be approximated using the normal distribution $\mathcal{N}$:
\begin{align}
  \bar{X}_{N-1} \sim \mathcal{N}(\mu, \frac{\sigma^2}{N-1}), \label{eq:clt}
\end{align}
where $\mu$ and  $\sigma^2$ are the mean and variance of $X$, respectively.
Because of the asymmetry of $Y_{ij} = - Y_{ji}$, the mean $\mu$ is equal to zero.
When the mean is known \textit{a priori}, variance $\sigma^2$ of $X$ is given by
\begin{align}
  \sigma^2 = \frac{1}{N(N-1)}\sum_{i} \sum_{j \neq i} Y_{ij}^2. \label{eq:sample_variance}
\end{align}
Finally, we obtain the distribution of the potential $P(\hat{s})$:
\begin{align}
  \hat{s} \sim \mathcal{N}\left(0, \sigma_{\hat{s}} \right), \label{eq:potential_null_dist}
\end{align}
where $\sigma^2_{\hat{s}} = \frac{1}{N^3} \sum_{i} \sum_{j \neq i} Y_{ij}^2 $.

\clearpage
\section{$p$-value of each geographical zone}
\begin{figure}[htb]
  \includegraphics[width=\linewidth]{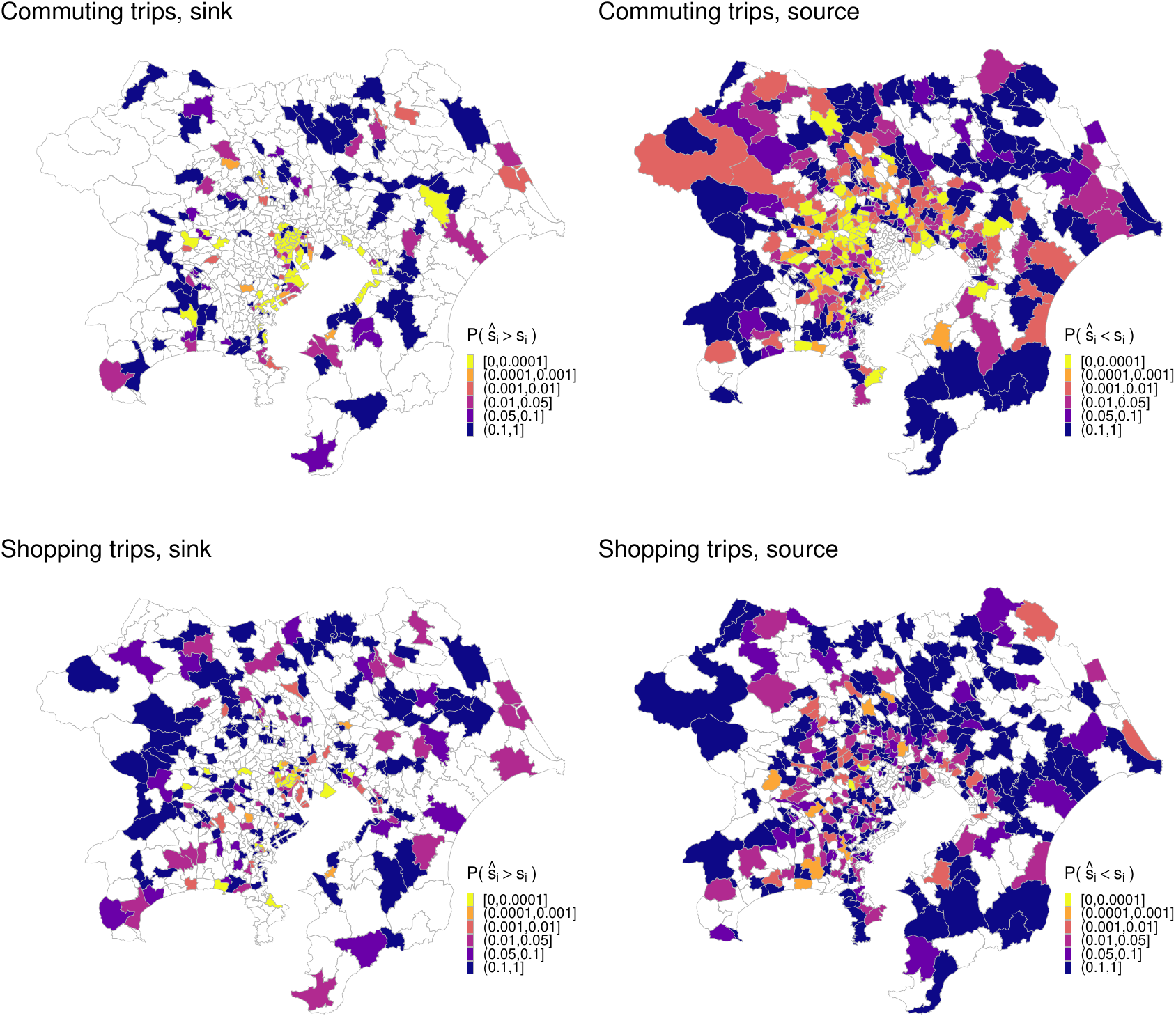}
  \caption{
$p$-value of each geographical zone for commuting and shopping trips.
}
  \label{fig:pvalue}
\end{figure}

Figure \ref{fig:pvalue} shows the $p$-value of each zone.
As described in the Methods section, for zones with positive potential, a multiple-comparison test is performed using the $p$-value of $P(\hat{s}_i > s_i)$ under the control of the false discovery rate (FDR).
For zones with negative potential, 
the test was using the $p$-value of $P(\hat{s}_i < s_i)$ under FDR control.

\clearpage
\section{Comparison with previous methods}
Here, we describe the details of the previous method applied to the trip datasets in Tokyo metropolitan area in the ``Comparison with previous methods'' section in the main text.

\subsection*{Incoming trips} 
According to \cite{Burger2012}, the definition of incoming trips is given by
\begin{align}
  \text{Incoming trips at zone $i$: $D^{in}_i$} = \sum_{j \neq i} M_{ji},
\end{align}
where $M$ is the OD matrix whose element $M_{ij}$ indicates the number of trips from the $i$th zone to the $j$th zone.

We also define the density of incoming trips in zone $i$ as
$$
d^{in}_i = D^{in}_i/ A_i,
$$
where $A_i$ denotes the area of $i$th zone.

\subsection*{Clustering method} 
In the seminal paper of \citet{GIULIANO1991}, a center is defined as ``a contiguous set of zones, each with density above some cutoff $\bar{D}$, that together have at least $\bar{E}$ total employment and for which all the immediately adjacent zones outside the subcenter have a density below $\bar{D}$.''
We applied this method to  spatial distribution of incoming trip density $d^{in}_i$.
We adopted the cutoff parameters proposed by \citet{MunzGarcia2010} as follows:
Density threshold $\bar{D}$ is the average density of the total area
and the threshold for the total incoming trips within a contiguous set is 1\% of the total incoming trips in the Tokyo metropolitan area.

\subsection*{LISA method} 
According to  previous studies \citep{Arribas-Bel2015,Li2018},  local Moran's $I_i$ is given by
\begin{align}
  I_i = z_i \sum_j w_{ij} z_j,
\end{align}
where $z$ is the z-score, $z_i = \frac{d^{in}_i - \mu}{\sigma}$. $\mu,\sigma$ are the mean and standard deviation of $d^{in}$, respectively.
$w_{ij}$ is the weighted matrix, given by
$$
w_{ij} =
\begin{cases}
1 \quad (\text{$i$ and $j$ are adjacent})\\
0 \quad (\text{otherwise})
\end{cases}.
$$

The previous studies have introduced a null model with random permutation to evaluate the statistical significance of the obtained $I_i$.
Given a randomly permuted $\hat{z}_i$, the null model of the Moran $I$ index is given by
\begin{align}
  \hat{I}_i = z_i \sum_j w_{ij} \hat{z}_j,
\end{align}
The empirical distribution of $\hat{I}_i$ is obtained from 99,999 observations, and the pseudo p-values are obtained by $P(\hat{I}_i > I_i)$.
We performed a multiple-comparison test under the control of the false discovery rate (FDR), $\alpha$ = 0.05, using a standard Benjamini-Hochberg procedure \citep{BHMethod}.

According to Moran's scatterplot \citep{Anselin1996}, the zones with significant $I_i$ are classified into four types, and two of them are defined as city centers: high-high zones (high density with high-density neighbors) and high-low zones (high density with low-density neighbors).

\subsection*{The method proposed by \citet{Veneri2013}} 
\citet{Veneri2013} defined the ratio between incoming and outgoing trips as flow centrality ($FC$) and proposed a necessary condition of the center by $FC_i >1$. 
This condition is written as
\begin{align}
 FC_i \equiv \frac{\sum_{j=1}^N M_{ji}}{\sum_{j=1}^N M_{ij}} > 1
 \Leftrightarrow  \sum_{j \neq i }^N M_{ji} -  \sum_{j \neq i}^N M_{ij}  > 0.
\end{align}
This is equivalent to the condition that the potential in Equation (6) is positive, $s_i > 0$.

Under this method, an additional condition by directional dominance index ($DII$) is imposed:
\begin{align}
  DII_i \equiv \frac{D^{in}_i}{\frac{1}{N}\sum_j D^{in}_{j}} > 1
\Leftrightarrow  D^{in}_i > \frac{1}{N}\sum_j  D^{in}_j.
\end{align}
This condition indicates that the incoming trips are above the mean.

\citet{Veneri2013} defined the zones that satisfy both conditions as ``second-order subcenters.''

\clearpage

\hypertarget{tutorial-potential-of-human-flow-general-graph-case}{%
\section{Tutorial: Potential of Human flow (general graph
case)}\label{tutorial-potential-of-human-flow-general-graph-case}}

This tutorial demonstrates how to detect significant sinks and sources
of human flow according to the scalar potential in the case of the Tokyo
metropolitan area.

\hypertarget{dataset-origin-destination-matrix-in-tokyo-with-trip-distance}{%
\subsection{Dataset: origin-destination matrix in Tokyo with trip
distance}\label{dataset-origin-destination-matrix-in-tokyo-with-trip-distance}}

We use a combined dataset of commuting and shopping trips from the person-trip surveys conducted in 2018 in the Tokyo metropolitan area (see the Data section in main text for details).

\begin{Shaded}
\begin{Highlighting}[]
\FunctionTok{suppressPackageStartupMessages}\NormalTok{(}\FunctionTok{library}\NormalTok{(tidyverse))}
\NormalTok{od }\OtherTok{=} \FunctionTok{read\_csv}\NormalTok{(}\StringTok{"tokyo2018.csv"}\NormalTok{,}
              \AttributeTok{col\_types=}\FunctionTok{list}\NormalTok{(}\AttributeTok{trips =} \StringTok{"n"}\NormalTok{, }\AttributeTok{road\_distance\_in\_km=}\StringTok{"n"}\NormalTok{,}\AttributeTok{.default=}\StringTok{"c"}\NormalTok{)) }
\NormalTok{od}
\CommentTok{\#\textgreater{} \# A tibble: 377,610 x 4}
\CommentTok{\#\textgreater{}    origin dest  trips road\_distance\_in\_km}
\CommentTok{\#\textgreater{}    \textless{}chr\textgreater{}  \textless{}chr\textgreater{} \textless{}dbl\textgreater{}               \textless{}dbl\textgreater{}}
\CommentTok{\#\textgreater{}  1 0011   0010    990                3.24}
\CommentTok{\#\textgreater{}  2 0012   0010   2445                2.00}
\CommentTok{\#\textgreater{}  3 0020   0010   2793                2.81}
\CommentTok{\#\textgreater{}  4 0021   0010      0                2.07}
\CommentTok{\#\textgreater{}  5 0022   0010      0                2.26}
\CommentTok{\#\textgreater{}  6 0023   0010   1692                3.30}
\CommentTok{\#\textgreater{}  7 0024   0010   1318                4.25}
\CommentTok{\#\textgreater{}  8 0030   0010   2525                4.65}
\CommentTok{\#\textgreater{}  9 0031   0010    893                3.64}
\CommentTok{\#\textgreater{} 10 0032   0010   1426                8.81}
\CommentTok{\#\textgreater{} \# i 377,600 more rows}
\end{Highlighting}
\end{Shaded}

The first two columns indicate the origin and destination zones,
representing the geographical zone code defined by the Tokyo
Metropolitan Region Transportation Planning Commission. The \emph{trips}
column shows the number of trips. We added the
\emph{road\_distance\_in\_km} column showing the trip distance from the
origin to the destination. This distance is evaluated by the road
distance between the centroids of the origin and destination zones,
calculated by the \href{https://project-osrm.org/}{Open Source Routing
Machine (OSRM)} based on \href{https://www.openstreetmap.org/}{Open
Street Map}. The dataset contains 615 unique zones and 615 \(\times\)
614 rows (excluding the trips within each zone).

\hypertarget{preprocess-convert-the-given-dataset-to-net-flow-y-on-a-graph-gve}{%
\subsection{\texorpdfstring{Preprocess: convert the given dataset to
net-flow \(Y\) on a graph
\(G(V,E)\)}{Preprocess: convert the given dataset to net-flow Y on a graph G(V,E)}}\label{preprocess-convert-the-given-dataset-to-net-flow-y-on-a-graph-gve}}

As described in the main text, we denoted the net movements of people as
\(Y\) on a graph \$G(V,E). In the current context, the vertices \(V\)
correspond to the locations and the edges \(E\) are the connections
between them.

Distances \(d_{ij}\) between locations \(i\) and \(j\) crucially affect
their connections. Trips will be rare for long distances even if the
potentials at the endpoints differ. On the other hand, no trips are
often observed for short distances. This implies that zero movements
happen in two distinct situations. To distinguish the nonexistence of
movements based on distance and other factors, we assumed that a pair of
locations is connected if its road distance \(d_{ij}\) is within a
threshold \(\theta\).

We calculate the threshold distance \(\theta\) above which trips are
rarely observed. We show the cumulative distribution of trip distance
\(P(X<d)\) as a function of trip distance \(d\). The threshold
\(\theta\) is determined based on criterion \(P(X<\theta) = 0.99\).

\begin{Shaded}
\begin{Highlighting}[]
\NormalTok{df }\OtherTok{=}\NormalTok{ od }\SpecialCharTok{\%\textgreater{}\%} \FunctionTok{arrange}\NormalTok{(road\_distance\_in\_km) }\SpecialCharTok{\%\textgreater{}\%} \FunctionTok{mutate}\NormalTok{(}\AttributeTok{cum\_trips =} \FunctionTok{cumsum}\NormalTok{(trips))}
\NormalTok{df }\OtherTok{=}\NormalTok{ df }\SpecialCharTok{\%\textgreater{}\%} \FunctionTok{mutate}\NormalTok{( }\AttributeTok{cum\_dist =}\NormalTok{ cum\_trips }\SpecialCharTok{/} \FunctionTok{sum}\NormalTok{(df}\SpecialCharTok{$}\NormalTok{trips))}
\NormalTok{df}
\CommentTok{\#\textgreater{} \# A tibble: 377,610 x 6}
\CommentTok{\#\textgreater{}    origin dest  trips road\_distance\_in\_km cum\_trips  cum\_dist}
\CommentTok{\#\textgreater{}    \textless{}chr\textgreater{}  \textless{}chr\textgreater{} \textless{}dbl\textgreater{}               \textless{}dbl\textgreater{}     \textless{}dbl\textgreater{}     \textless{}dbl\textgreater{}}
\CommentTok{\#\textgreater{}  1 1021   1020    208                1.00       208 0.0000133}
\CommentTok{\#\textgreater{}  2 1020   1021   1643                1.00      1851 0.000118 }
\CommentTok{\#\textgreater{}  3 1022   1021    640                1.04      2491 0.000159 }
\CommentTok{\#\textgreater{}  4 1021   1022    904                1.04      3395 0.000216 }
\CommentTok{\#\textgreater{}  5 3017   3013   1383                1.17      4778 0.000305 }
\CommentTok{\#\textgreater{}  6 3013   3017    829                1.17      5607 0.000358 }
\CommentTok{\#\textgreater{}  7 6023   6021   1820                1.17      7427 0.000474 }
\CommentTok{\#\textgreater{}  8 6021   6023   1035                1.17      8462 0.000540 }
\CommentTok{\#\textgreater{}  9 6061   6060   2601                1.18     11063 0.000705 }
\CommentTok{\#\textgreater{} 10 6060   6061    216                1.18     11279 0.000719 }
\CommentTok{\#\textgreater{} \# i 377,600 more rows}

\CommentTok{\# find intersection with threshold}
\NormalTok{y0 }\OtherTok{=} \FloatTok{0.99}
\NormalTok{RootLinearInterpolant }\OtherTok{\textless{}{-}} \ControlFlowTok{function}\NormalTok{ (x, y, }\AttributeTok{y0 =} \DecValTok{0}\NormalTok{) \{}
  \ControlFlowTok{if}\NormalTok{ (}\FunctionTok{is.unsorted}\NormalTok{(x)) \{}
\NormalTok{    ind }\OtherTok{\textless{}{-}} \FunctionTok{order}\NormalTok{(x)}
\NormalTok{    x }\OtherTok{\textless{}{-}}\NormalTok{ x[ind]; y }\OtherTok{\textless{}{-}}\NormalTok{ y[ind]}
\NormalTok{  \}}
\NormalTok{  z }\OtherTok{\textless{}{-}}\NormalTok{ y }\SpecialCharTok{{-}}\NormalTok{ y0}
\NormalTok{  k }\OtherTok{\textless{}{-}} \FunctionTok{which}\NormalTok{(z[}\SpecialCharTok{{-}}\DecValTok{1}\NormalTok{] }\SpecialCharTok{*}\NormalTok{ z[}\SpecialCharTok{{-}}\FunctionTok{length}\NormalTok{(z)] }\SpecialCharTok{\textless{}} \DecValTok{0}\NormalTok{)}
\NormalTok{  xk }\OtherTok{\textless{}{-}}\NormalTok{ x[k] }\SpecialCharTok{{-}}\NormalTok{ z[k] }\SpecialCharTok{*}\NormalTok{ (x[k }\SpecialCharTok{+} \DecValTok{1}\NormalTok{] }\SpecialCharTok{{-}}\NormalTok{ x[k]) }\SpecialCharTok{/}\NormalTok{ (z[k }\SpecialCharTok{+} \DecValTok{1}\NormalTok{] }\SpecialCharTok{{-}}\NormalTok{ z[k])}
  \FunctionTok{return}\NormalTok{(xk)}
\NormalTok{\}}
\NormalTok{thres }\OtherTok{=} \FunctionTok{RootLinearInterpolant}\NormalTok{(df}\SpecialCharTok{$}\NormalTok{road\_distance\_in\_km, df}\SpecialCharTok{$}\NormalTok{cum\_dist, y0)}

\FunctionTok{library}\NormalTok{(ggplot2)}
\FunctionTok{theme\_set}\NormalTok{(}\FunctionTok{theme\_minimal}\NormalTok{())}
\FunctionTok{ggplot}\NormalTok{(df, }\FunctionTok{aes}\NormalTok{(}\AttributeTok{x =}\NormalTok{ road\_distance\_in\_km, }\AttributeTok{y=}\NormalTok{cum\_dist) ) }\SpecialCharTok{+}
  \FunctionTok{geom\_line}\NormalTok{(}\AttributeTok{linewidth=}\FloatTok{1.3}\NormalTok{) }\SpecialCharTok{+} 
  \FunctionTok{xlab}\NormalTok{(}\StringTok{"Distance [km]"}\NormalTok{) }\SpecialCharTok{+} 
  \FunctionTok{ylab}\NormalTok{(}\StringTok{"Cumulative distribution, Prob. (X \textless{} d)"}\NormalTok{)  }\SpecialCharTok{+} 
  \FunctionTok{geom\_hline}\NormalTok{(}\AttributeTok{yintercept=}\NormalTok{ y0, }\AttributeTok{color=}\StringTok{"red"}\NormalTok{, }\AttributeTok{linetype=}\StringTok{"dashed"}\NormalTok{) }\SpecialCharTok{+}
  \FunctionTok{annotate}\NormalTok{(}\StringTok{"segment"}\NormalTok{, }\AttributeTok{x=}\NormalTok{thres,}\AttributeTok{xend=}\NormalTok{thres, }\AttributeTok{y=}\NormalTok{y0 }\SpecialCharTok{{-}} \FloatTok{0.1}\NormalTok{, }\AttributeTok{yend=}\NormalTok{y0, }
           \AttributeTok{arrow=}\FunctionTok{arrow}\NormalTok{(}\AttributeTok{length =} \FunctionTok{unit}\NormalTok{(}\DecValTok{2}\NormalTok{, }\StringTok{"mm"}\NormalTok{)) ) }\SpecialCharTok{+} 
  \FunctionTok{annotate}\NormalTok{(}\StringTok{"text"}\NormalTok{, }\AttributeTok{x =}\NormalTok{ thres, }\AttributeTok{y =}\NormalTok{ y0 }\SpecialCharTok{{-}} \FloatTok{0.15}\NormalTok{, }
           \AttributeTok{label=}\FunctionTok{sprintf}\NormalTok{(}\StringTok{"\%.1fkm"}\NormalTok{, thres),}\AttributeTok{size=}\DecValTok{3}\NormalTok{)}
\end{Highlighting}
\end{Shaded}

\includegraphics[width=0.8\linewidth]{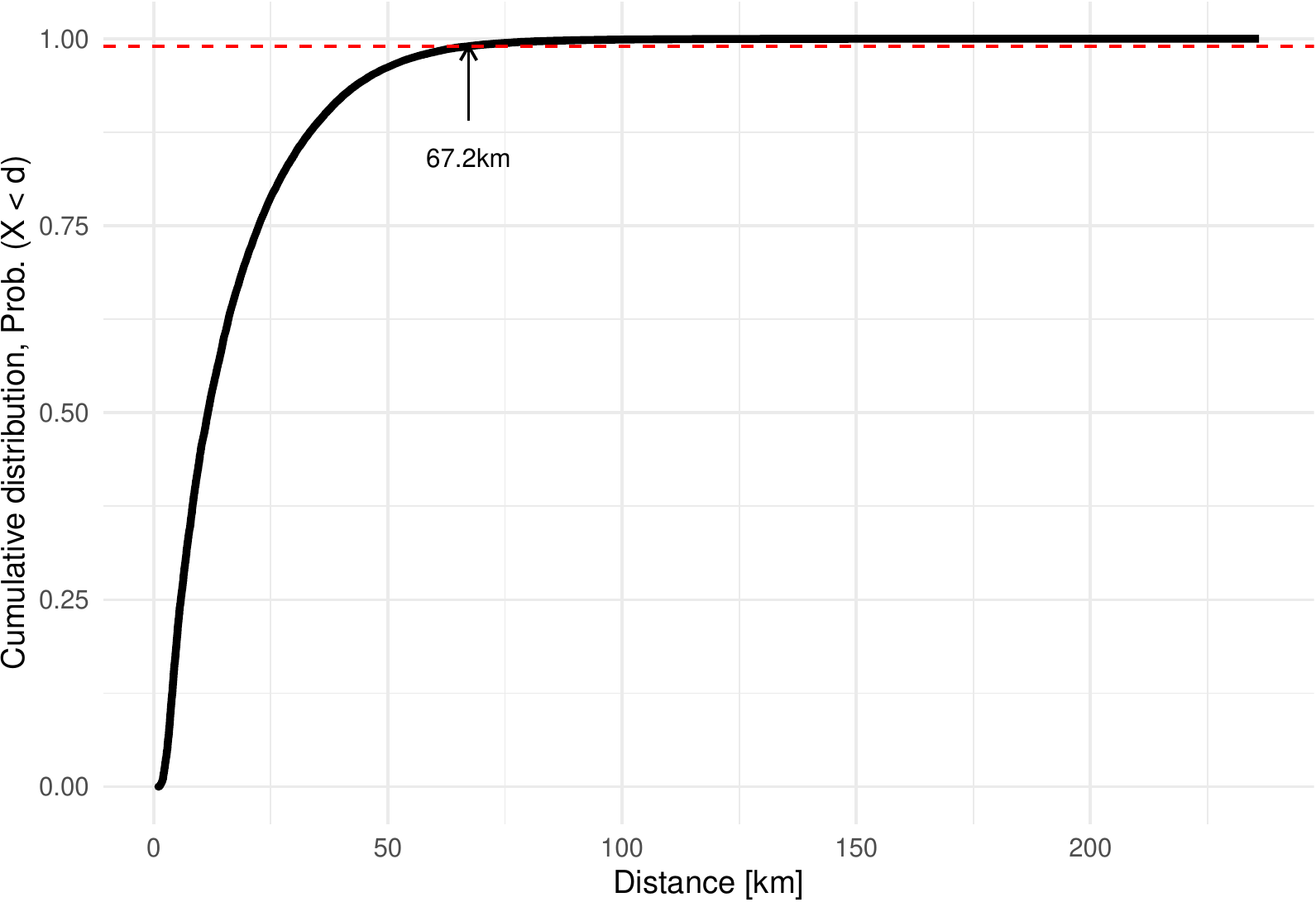}

Next, we make net flows between all pairs of locations and select them
by \(d_{ij} < \theta\).

\begin{Shaded}
\begin{Highlighting}[]
\NormalTok{od\_forward }\OtherTok{=}\NormalTok{ od }
\NormalTok{od\_backward }\OtherTok{=}\NormalTok{ od\_forward }\SpecialCharTok{\%\textgreater{}\%} \FunctionTok{select}\NormalTok{(origin,dest,trips) }\SpecialCharTok{\%\textgreater{}\%} 
  \FunctionTok{rename}\NormalTok{(}\AttributeTok{origin\_new =}\NormalTok{ dest, }\AttributeTok{dest\_new =}\NormalTok{ origin) }\SpecialCharTok{\%\textgreater{}\%} 
  \FunctionTok{rename}\NormalTok{(}\AttributeTok{origin =}\NormalTok{ origin\_new, }\AttributeTok{dest =}\NormalTok{ dest\_new, }\AttributeTok{back\_trips =}\NormalTok{ trips ) }\SpecialCharTok{\%\textgreater{}\%}
  \FunctionTok{select}\NormalTok{(origin, dest, back\_trips)}


\NormalTok{netflow }\OtherTok{=}\NormalTok{ od\_forward }\SpecialCharTok{\%\textgreater{}\%} \FunctionTok{left\_join}\NormalTok{(od\_backward,}\AttributeTok{by =} \FunctionTok{join\_by}\NormalTok{(origin, dest)) }\SpecialCharTok{\%\textgreater{}\%}
  \FunctionTok{filter}\NormalTok{(origin }\SpecialCharTok{\textgreater{}}\NormalTok{ dest) }\SpecialCharTok{\%\textgreater{}\%} \FunctionTok{mutate}\NormalTok{(}\AttributeTok{netflow =}\NormalTok{ trips }\SpecialCharTok{{-}}\NormalTok{ back\_trips) }\SpecialCharTok{\%\textgreater{}\%} 
  \FunctionTok{rename}\NormalTok{(}\AttributeTok{vertex1 =}\NormalTok{ origin, }\AttributeTok{vertex2=}\NormalTok{dest)}

\NormalTok{netflow\_selected }\OtherTok{=}\NormalTok{ netflow }\SpecialCharTok{\%\textgreater{}\%} 
  \FunctionTok{filter}\NormalTok{(road\_distance\_in\_km }\SpecialCharTok{\textless{}}\NormalTok{ thres) }\SpecialCharTok{\%\textgreater{}\%} 
  \FunctionTok{select}\NormalTok{(vertex1,vertex2,netflow)}

\NormalTok{netflow\_selected}
\CommentTok{\#\textgreater{} \# A tibble: 117,329 x 3}
\CommentTok{\#\textgreater{}    vertex1 vertex2 netflow}
\CommentTok{\#\textgreater{}    \textless{}chr\textgreater{}   \textless{}chr\textgreater{}     \textless{}dbl\textgreater{}}
\CommentTok{\#\textgreater{}  1 0011    0010        990}
\CommentTok{\#\textgreater{}  2 0012    0010       2323}
\CommentTok{\#\textgreater{}  3 0020    0010       2793}
\CommentTok{\#\textgreater{}  4 0021    0010          0}
\CommentTok{\#\textgreater{}  5 0022    0010          0}
\CommentTok{\#\textgreater{}  6 0023    0010       1692}
\CommentTok{\#\textgreater{}  7 0024    0010       1318}
\CommentTok{\#\textgreater{}  8 0030    0010       2525}
\CommentTok{\#\textgreater{}  9 0031    0010        893}
\CommentTok{\#\textgreater{} 10 0032    0010       1426}
\CommentTok{\#\textgreater{} \# i 117,319 more rows}
\end{Highlighting}
\end{Shaded}

\hypertarget{calculate-the-potential-s-for-the-net-flow-y}{%
\subsection{\texorpdfstring{Calculate the potential \(s\) for the net
flow
\(Y\)}{Calculate the potential s for the net flow Y}}\label{calculate-the-potential-s-for-the-net-flow-y}}

The potential in Equation (1) in the main text is calculated by using
function \emph{scalar\_potential\_on\_graph} in the
\emph{HodgePotentialHumanFlow} package.

\begin{Shaded}
\begin{Highlighting}[]
\FunctionTok{library}\NormalTok{(HodgePotentialHumanFlow)}
\NormalTok{s }\OtherTok{=}\NormalTok{ HodgePotentialHumanFlow}\SpecialCharTok{::}\FunctionTok{scalar\_potential\_on\_graph}\NormalTok{(netflow\_selected)}
\CommentTok{\#\textgreater{} Calculating p{-}values by monte calro (100000 steps).}
\FunctionTok{head}\NormalTok{(s)}
\CommentTok{\#\textgreater{}   zone potential pvalue\_above}
\CommentTok{\#\textgreater{} 1 0011  406.7119            0}
\CommentTok{\#\textgreater{} 2 0012  494.1058            0}
\CommentTok{\#\textgreater{} 3 0020  261.1093            0}
\CommentTok{\#\textgreater{} 4 0021  222.9379            0}
\CommentTok{\#\textgreater{} 5 0022  191.7659            0}
\CommentTok{\#\textgreater{} 6 0023  227.1667            0}
\end{Highlighting}
\end{Shaded}

The function returns a dataframe with three columns: zone, potential,
and \(p\)-value. The \(p\)-value indicates \(P(\hat{s}_i > s_i)\) under
the null hypothesis (see the main text for details). The sample number
for calculating this pseudo \(p\)-value can be specified by option
\emph{num\_samples} (default = 1e5).

\begin{Shaded}
\begin{Highlighting}[]
\NormalTok{s }\OtherTok{=}\NormalTok{ HodgePotentialHumanFlow}\SpecialCharTok{::}\FunctionTok{scalar\_potential\_on\_graph}\NormalTok{(netflow\_selected, }\AttributeTok{num\_samples =} \FloatTok{1e6}\NormalTok{)}
\CommentTok{\#\textgreater{} Calculating p{-}values by monte calro (1000000 steps).}
\end{Highlighting}
\end{Shaded}

We plot the potential on a map using a shapefile publicly available from
\href{https://www.tokyo-pt.jp/data/01_01}{the Tokyo Metropolitan Region
Transportation Planning Commission}.

\begin{Shaded}
\begin{Highlighting}[]
\FunctionTok{library}\NormalTok{(sf)}
\CommentTok{\#\textgreater{} Linking to GEOS 3.11.1, GDAL 3.6.3, PROJ 9.1.1; sf\_use\_s2() is TRUE}
\NormalTok{map }\OtherTok{=} \FunctionTok{st\_read}\NormalTok{(}\StringTok{"tokyo2018.gpkg"}\NormalTok{, }\AttributeTok{quiet=}\NormalTok{T) }
\NormalTok{map }\OtherTok{=}\NormalTok{ map }\SpecialCharTok{\%\textgreater{}\%} \FunctionTok{left\_join}\NormalTok{(s)}
\CommentTok{\#\textgreater{} Joining with \textasciigrave{}by = join\_by(zone)\textasciigrave{}}

\FunctionTok{library}\NormalTok{(ggplot2)}
\FunctionTok{library}\NormalTok{(viridis,}\AttributeTok{quietly =}\NormalTok{ T)}
\FunctionTok{suppressPackageStartupMessages}\NormalTok{(}\FunctionTok{library}\NormalTok{(urbnthemes,}\AttributeTok{quietly=}\NormalTok{T))}
\FunctionTok{library}\NormalTok{(ggspatial,}\AttributeTok{quietly=}\NormalTok{T)}
\FunctionTok{suppressPackageStartupMessages}\NormalTok{(}\FunctionTok{library}\NormalTok{(scales,}\AttributeTok{quiet =}\NormalTok{ T))}


\FunctionTok{ggplot}\NormalTok{(}\AttributeTok{data=}\NormalTok{map, }\FunctionTok{aes}\NormalTok{(}\AttributeTok{fill=}\NormalTok{potential)) }\SpecialCharTok{+} 
  \FunctionTok{geom\_sf}\NormalTok{(}\AttributeTok{color=}\StringTok{"black"}\NormalTok{, }\AttributeTok{lwd=}\FloatTok{0.2}\NormalTok{)  }\SpecialCharTok{+}
  \FunctionTok{scale\_fill\_viridis\_c}\NormalTok{(}\AttributeTok{option=}\StringTok{"viridis"}\NormalTok{, }\AttributeTok{trans=}\FunctionTok{modulus\_trans}\NormalTok{(}\FloatTok{0.3}\NormalTok{)) }\SpecialCharTok{+}
  \FunctionTok{theme\_urbn\_map}\NormalTok{(}\AttributeTok{base\_family=}\StringTok{"Helvetica"}\NormalTok{) }\SpecialCharTok{+}
  \FunctionTok{theme}\NormalTok{(}\AttributeTok{legend.key.height=}\FunctionTok{unit}\NormalTok{(}\FloatTok{1.5}\NormalTok{,}\StringTok{"line"}\NormalTok{)) }\SpecialCharTok{+}
  \FunctionTok{annotation\_scale}\NormalTok{(}\AttributeTok{location =} \StringTok{"bl"}\NormalTok{) }\SpecialCharTok{+} 
  \FunctionTok{annotation\_north\_arrow}\NormalTok{(}\AttributeTok{location =} \StringTok{"br"}\NormalTok{, }\AttributeTok{which\_north =} \StringTok{"grid"}\NormalTok{, }
  \AttributeTok{style =} \FunctionTok{north\_arrow\_fancy\_orienteering}\NormalTok{(}\AttributeTok{text\_size =}\DecValTok{0}\NormalTok{, }\AttributeTok{line\_width=}\FloatTok{0.5}\NormalTok{)) }
\end{Highlighting}
\end{Shaded}

\includegraphics[width=0.8\linewidth]{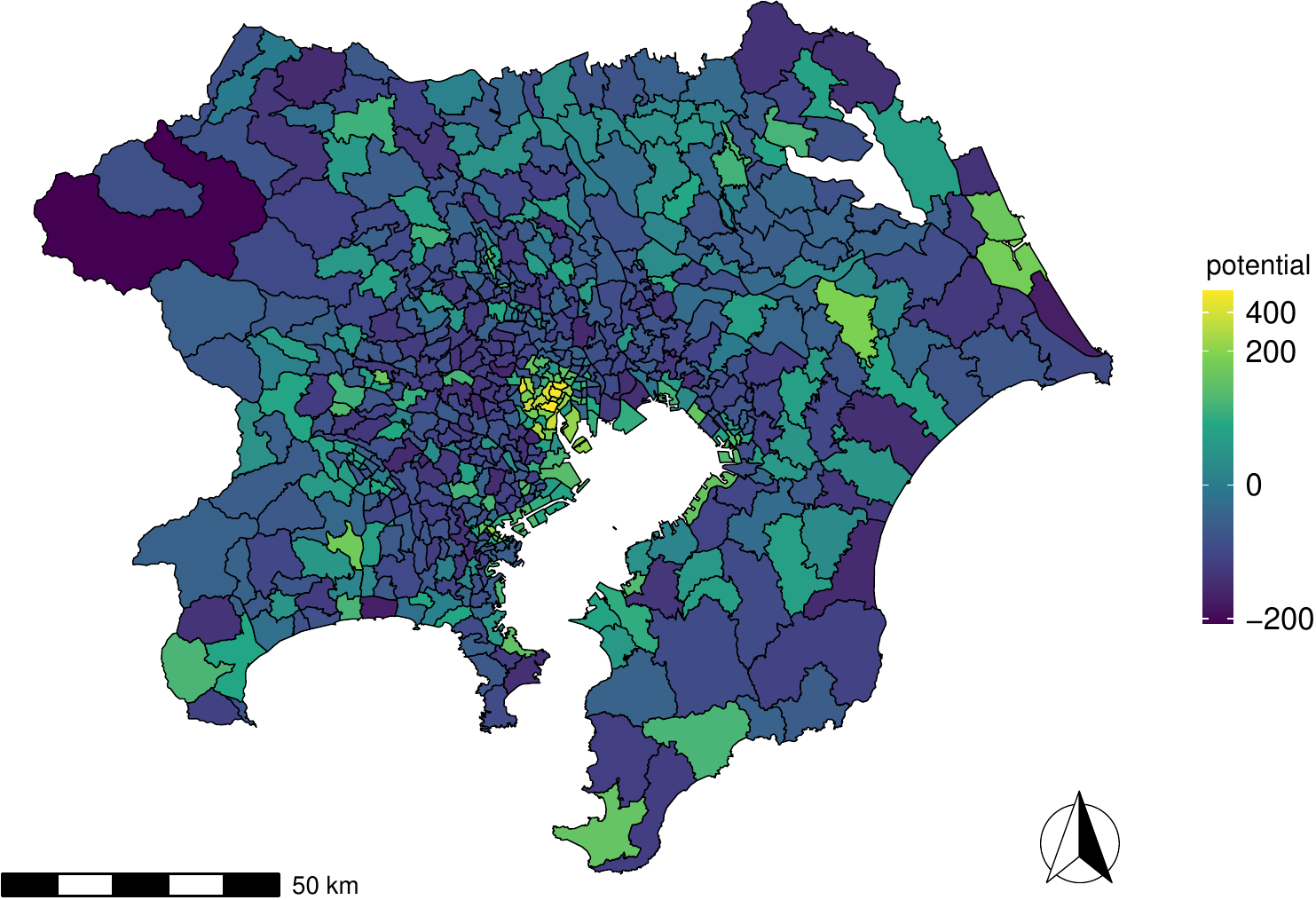}

\hypertarget{detect-significant-sinks-and-sources}{%
\subsection{Detect significant sinks and
sources}\label{detect-significant-sinks-and-sources}}

To detect significant sinks for zones with a positive potential, a
multiple-comparison test is performed using the \(p\)-value under the
control of the false discovery rate (FDR), \(\alpha (=0.05)\), with a
standard Benjamini-Hochberg procedure. Similarly, for zones with a
negative potential, a multiple-comparison test is performed using the
\(p\)-value of \(1 - P(\hat{s}_i > s_i)\) under FDR control.

\begin{Shaded}
\begin{Highlighting}[]
\NormalTok{df\_sink}\OtherTok{=}\NormalTok{ s }\SpecialCharTok{\%\textgreater{}\%} \FunctionTok{filter}\NormalTok{(potential }\SpecialCharTok{\textgreater{}=} \DecValTok{0}\NormalTok{)}
\NormalTok{df\_sink}\SpecialCharTok{$}\NormalTok{pvalue\_adj }\OtherTok{=} \FunctionTok{p.adjust}\NormalTok{(df\_sink}\SpecialCharTok{$}\NormalTok{pvalue\_above, }\AttributeTok{method=}\StringTok{"fdr"}\NormalTok{)}
\NormalTok{df\_source}\OtherTok{=}\NormalTok{ s }\SpecialCharTok{\%\textgreater{}\%} \FunctionTok{filter}\NormalTok{(potential }\SpecialCharTok{\textless{}} \DecValTok{0}\NormalTok{)}
\NormalTok{df\_source}\SpecialCharTok{$}\NormalTok{pvalue\_adj }\OtherTok{=} \FunctionTok{p.adjust}\NormalTok{(}\DecValTok{1} \SpecialCharTok{{-}}\NormalTok{ df\_source}\SpecialCharTok{$}\NormalTok{pvalue\_above, }\AttributeTok{method=}\StringTok{"fdr"}\NormalTok{)}
\NormalTok{df }\OtherTok{=} \FunctionTok{rbind}\NormalTok{(df\_sink, df\_source)}

\FunctionTok{library}\NormalTok{(sf)}
\NormalTok{alpha }\OtherTok{=} \FloatTok{0.05}
\NormalTok{map }\OtherTok{=} \FunctionTok{st\_read}\NormalTok{(}\StringTok{"tokyo2018.gpkg"}\NormalTok{, }\AttributeTok{quiet=}\NormalTok{T) }
\NormalTok{map }\OtherTok{=}\NormalTok{ map }\SpecialCharTok{\%\textgreater{}\%} \FunctionTok{left\_join}\NormalTok{(df) }\SpecialCharTok{\%\textgreater{}\%} \FunctionTok{rowwise}\NormalTok{() }\SpecialCharTok{\%\textgreater{}\%} 
   \FunctionTok{mutate}\NormalTok{(}\AttributeTok{type =} \FunctionTok{case\_when}\NormalTok{(}
\NormalTok{                      potential }\SpecialCharTok{\textgreater{}=} \DecValTok{0} \SpecialCharTok{\&}\NormalTok{ pvalue\_adj }\SpecialCharTok{\textless{}}\NormalTok{ alpha }\SpecialCharTok{\textasciitilde{}} \StringTok{"Significant sink"}\NormalTok{,}
\NormalTok{                      potential }\SpecialCharTok{\textless{}} \DecValTok{0} \SpecialCharTok{\&}\NormalTok{ pvalue\_adj }\SpecialCharTok{\textless{}}\NormalTok{ alpha }\SpecialCharTok{\textasciitilde{}} \StringTok{"Significant source"}\NormalTok{,}
\NormalTok{   ))}
\CommentTok{\#\textgreater{} Joining with \textasciigrave{}by = join\_by(zone)\textasciigrave{}}
\FunctionTok{ggplot}\NormalTok{(}\AttributeTok{data=}\NormalTok{map, }\FunctionTok{aes}\NormalTok{(}\AttributeTok{fill=}\NormalTok{type)) }\SpecialCharTok{+} 
  \FunctionTok{geom\_sf}\NormalTok{(}\AttributeTok{color=}\StringTok{"black"}\NormalTok{, }\AttributeTok{lwd=}\FloatTok{0.2}\NormalTok{)  }\SpecialCharTok{+}
  \FunctionTok{theme\_urbn\_map}\NormalTok{(}\AttributeTok{base\_family=}\StringTok{"Helvetica"}\NormalTok{) }\SpecialCharTok{+}
  \FunctionTok{scale\_fill\_manual}\NormalTok{(}\AttributeTok{values =} \FunctionTok{c}\NormalTok{(}\StringTok{"\#e41a1c"}\NormalTok{,}\StringTok{"navyblue"}\NormalTok{),}\AttributeTok{na.translate =}\NormalTok{ F) }\SpecialCharTok{+}
  \FunctionTok{theme}\NormalTok{(}\AttributeTok{legend.title=}\FunctionTok{element\_blank}\NormalTok{())}
\end{Highlighting}
\end{Shaded}

\includegraphics[width=0.8\linewidth]{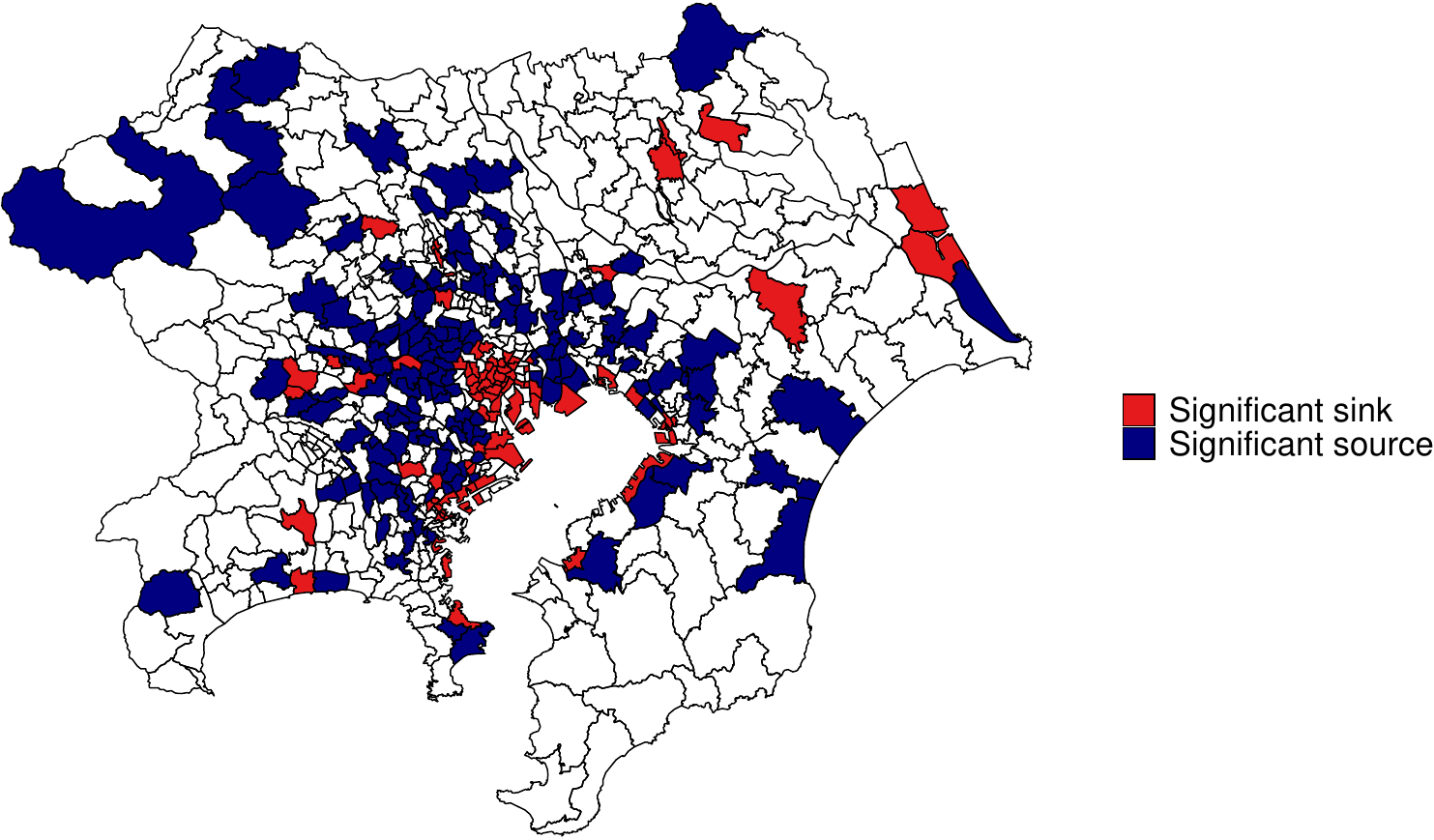}

%
